\documentclass{article} 
\usepackage{nips13submit_e,times}
\usepackage{hyperref}
\usepackage{url}

\usepackage{titlesec}
\usepackage[ruled,vlined]{algorithm2e}
\usepackage{amsmath}
\usepackage{graphicx}
\usepackage{url}
\usepackage{natbib}
\usepackage[noabbrev,capitalise]{cleveref}
\usepackage{lipsum}
\usepackage{amssymb}
\usepackage{multicol}
\usepackage{bm}
\usepackage[justification=centering]{caption}
\usepackage{wrapfig}
\usepackage{tikz}
\usepackage{float}
\usepackage{subcaption}
\usepackage{mathtools}
\usepackage{amsfonts}
\usepackage{bbm}

\usepackage{algpseudocode}

\usepackage{amsthm}
\usepackage{verbatim}

\DeclareMathOperator*{\argmax}{arg\,max}

\usepackage{xcolor}
\definecolor{forestgreen}{rgb}{0.13, 0.55, 0.13}

\usepackage{soul}

\usepackage{setspace}
\title{Co-Active Subspace Methods for the Joint Analysis of Adjacent Computer Models}

\author{Kellin N. Rumsey$^1$ \\
Statistical Sciences\\
Los Alamos National Laboratory \\
Los Alamos, NM 87545 \\
\texttt{knrumsey@lanl.gov} \\
\And
Zachary K. Hardy \\
X Theoretical Design Division \\
Los Alamos National Laboratory \\
Los Alamos, NM 87545 \\
\texttt{zhardy@lanl.gov} \\
\And
Cory Ahrens \\
X Theoretical Design Division \\
Los Alamos National Laboratory \\
Los Alamos, NM 87545 \\
\texttt{cdahrens@lanl.gov} \\
\And
Scott Vander Wiel \\
Statistical Sciences\\
Los Alamos National Laboratory \\
Los Alamos, NM 87545 \\
\texttt{scottv@lanl.gov}
}

\usepackage[normalem]{ulem}
\nipsfinalcopy 

\begin{document}

\maketitle

\normalsize

\doublespacing
\vspace*{-30pt}

\begin{abstract}
Active subspace (AS) methods are a valuable tool for understanding the relationship between the inputs and outputs of a Physics simulation. In this paper, an elegant generalization of the traditional ASM is developed to assess the co-activity of two computer models. This generalization, which we refer to as a Co-Active Subspace (Co-AS) Method, allows for the joint analysis of two or more computer models allowing for thorough exploration of the alignment (or non-alignment) of the respective gradient spaces. We define co-active directions, co-sensitivity indices, and a scalar ``concordance" metric (and complementary ``discordance" pseudo-metric) and we demonstrate that these are powerful tools for understanding the behavior of a class of computer models, especially when used to supplement traditional AS analysis. Details for efficient estimation of the Co-AS and an accompanying R package (\texttt{concordance}) are provided. Practical application is demonstrated through analyzing a set of simulated rate stick experiments for PBX 9501, a high explosive, offering insights into complex model dynamics.

\end{abstract}

\newpage

\standardsize

\section{Introduction}

In modern physical science, complex real-world systems are often modeled based on the current understanding of the physics governing the problem. These computer models (also called simulators) have proven to be valuable tools for advancing scientific knowledge. It is a common practice to consider a class of two or more computer models that endeavor to simulate the same (or similar) physical phenomena; we refer to these computer models as {\it adjacent}. Examples include: (i) computer models calibrated via distinct methods and/or data sources \citep{kennedy2001bayesian, michelsen2007modeling}, (ii) computer models approximated with different fidelities or emulated using different methods \citep{peherstorfer2018survey, zhang2020review, gramacy2020surrogates}, (iii) computer models grounded in competing theories of the underlying physics \citep{bernstein2019comparison, finnegan2022narrative, yannotty2023model}, and (iv) computer models describing neighboring physical processes. Within this context, the central focus revolves around comprehending the discrepancies and commonalities among these models, determining the point at which their similarity deems them as neighbors. We also seek to understand how the common inputs jointly affect these simulators.

For a single function (computer model) of interest, active subspace methods (ASM) \citep{constantine2015activebook} have emerged as a popular tool for exploring the intricate relationship between inputs and outputs within a physics simulator \citep{lukaczyk2014active, constantine2015discovering, constantine2017time, seshadri2018turbomachinery, tezzele2018combined, ji2019quantifying, batta2021uncovering}. ASM involves identifying a set of orthogonal directions within the input space that produce the largest changes in the response. In the case of multiple computer models (or equivalently, a vector-valued function), \cite{zahm2020gradient} and \cite{ji2018shared} propose methods for discovering a so-called {\it shared} active subspace. These approaches focus on finding a single subspace upon which all functions are simultaneously active. Commonly, the approach of \cite{zahm2020gradient} reduces to finding a key matrix $\bm C_k$ for each function of interest and working with their sum, $\bm H = \sum_{k=1}^K\bm C_k$. This approach has proven to be highly successful in theory and in practice.

In addition to active subspace based methods, the literature on multivariate sufficient dimension reduction (MSDR) is relevant to this discussion \citep{hsing1999nearest,setodji2004k, li2008projective,cook2022slice}. SDR methods, like sliced inverse regression \citep{li1991sliced}, attempt to find subspaces such that the conditional distribution of the model output given the inputs varies only within the subspace. The smallest such subspace is called the ``central subspace" and is the target of SDR. The central subspace can be challenging to find, and most methods rely on easier to find yet related subspaces (e.g., the central mean subspace). MSDR methods generalize these approaches to multiple functions (see \citep{dong2023selective} for a review).

Although closely related, the approach outlined in this manuscript differs from previous work in several important ways. First of all, our primary interest is not in finding a single active subspace which is effective for all functions (although it can be useful here). Rather, we are interested in understanding the complex relationship between two adjacent computer models and determining the directions within the input space along which each model's gradients are well-aligned. In particular, we introduce co-active directions, co-sensitivity indices, and a metric called \textit{concordance} that supplies a univariate measure of the alignment between the gradient spaces of two functions. So while previous work focuses on studying a class of $K$ functions, our methods are tailor-made for the pairwise study of two functions at a time. 

The motivation for this extension stemmed from the real-world challenge of leveraging existing data from one physical system to aid in the design of a similar, but not identical, new system. Standard engineering practice dictates that comprehensive full-system tests are essential to ensure the functionality of a new design \cite{buede2016engineering}. However, the unique context of our problem precludes such exhaustive testing, necessitating alternative means of design assurance. This is where the concept of co-activity comes into play.

In \cref{sec:background}, we provide a mathematical description of the problem and state several key questions which are of scientific interest. We also motivate the problem by describing rate stick experiment simulations with PBX 9501, a commonly studied high explosive (HE), and provide a brief review of traditional AS methods. Mathematical details and definitions for the Co-Active Subspace (Co-AS) method is given in \cref{sec:CASM}, and is illustrated using a simple class of multinomial computer models. We also provide details for efficiently estimating the Co-AS and refer the reader to the accompanying R package (available at \url{github.com/knrumsey/concordance}). \Cref{sec:HE} gives a sample analysis using the rate stick experiment simulations on PBX 9501, and concluding remarks are provided in \cref{sec:conclusions}. 

\section{Background and Motivation}
\label{sec:background}
\subsection{Adjacent Computer Models}
\label{sec:adjacent}
If computer models $f_1$ and $f_2$ share a common set of inputs $\bm x \in \mathcal X \subset \mathbb R^p$ and they describe the same (or similar) physical process, then we say that $f_1$ and $f_2$ are {\it adjacent} computer models. Any prior information over the inputs $\bm x$, represented by the density $\pi(\bm x)$, must be shared by all functions. The second condition, that the functions must describe the same process, is difficult to pin down precisely. To clarify this concept, we give three concrete examples. 
\begin{itemize}
    \item {\it Models with Different Fidelity or Resolution:} Consider a computer model which is characterized by a partial differential equation (PDE) with uncertain inputs $\bm x$. Often, this PDE must be solved numerically, such as via finite differences. Representing a highly accuracy solution, $f_1(\bm x)$, is solved on a high-resolution grid and may be expensive to run. Alternatively, $f_2(\bm x)$ is a ``low-fidelity" solution if it is solved over a coarse grid, perhaps sacrificing some accuracy for faster computation time \citep{peherstorfer2018survey}.
    \item {\it Differing Physics:} Variations of a computer model can arise when there are competing theories of the physics which govern the underlying process. For instance, \cite{bernstein2019comparison} consider a large class of computer models which attempt to predict material flow strength. The models differ only in the form and parameters of a material strength model. The relative accuracy of each model may depend on the material being modeled or on the region of input-space (i.e. pressure/temperature) of interest. 
    \item {\it Neighboring Systems:} Rather than modeling the exact same system, some model classes may attempt to explain/predict similar systems. In this scenario, one would seek to gain insight into an ``un-trusted'' model by comparing its behaviors to a ``trusted'' model, potentially validated against experimental data. A detailed example of this case is given in \cref{sec:pbxintro}. 
\end{itemize}

We are often interested in studying a class models, denoted $\mathcal F = \{f_1, f_2, \ldots, f_K\}$, in which all members of the class are pairwise adjacent. In these settings, there are a number of important questions a practitioner may seek to answer, such as
\begin{itemize}
    \item[Q1.] How similar are models $f_k$ and $f_\ell$ with respect to the inputs $\bm x$? In particular, do they respond similarly to perturbations? That is, are their active subspaces well-aligned?
    \item[Q2.] Can we partition (or cluster) the functions into groups of ``neighboring" models?
    \item[Q3.] Which inputs are the most ``important" simultaneously for functions $f_k$ and $f_\ell$? How does this compare to the effect of these inputs on each function independently?
    \item[Q4.] Can we produce a low dimensional representation of $\bm x$ that can be used to capture variability in both $f_k$ and $f_\ell$ simultaneously?
\end{itemize}
In this manuscript, we develop and present approaches which seek to answer these questions effectively. Q1 is discussed in \cref{sec:concordance}, where \cref{eq:conc} defines a scalar-value measure called {\it concordance}. Q2 is closely related to Q1 and our proposed solution (using the notion of {\it discordance}, \cref{eq:discord}) is demonstrated in \cref{fig:voronoi}.  We define {\it co-activity scores} in \cref{eq:signed,eq:unsigned} which can be used (in addition the {\it contribution vectors}, \cref{eq:contrib}) to address Q3, as demonstrated for PBX9501 in \cref{sec:cosenspbx}. Finally, Q4 is not the primary focus of this work, but it is an important question which many other authors have addressed. {\it Co-active directions} are defined in \cref{eq:coactivedir} and can sometimes be useful for this purpose. An example of this approach, along with a comparison to the method of \cite{zahm2020gradient}, is given in \cref{sec:lowD_JWL}.

\subsection{Rate Stick Experiments for PBX 9501}
\label{sec:pbxintro}
\begin{figure}[t!]
    \centering
    \begin{subfigure}[t]{0.48\textwidth}
        \centering
        \includegraphics[width=0.9\linewidth]{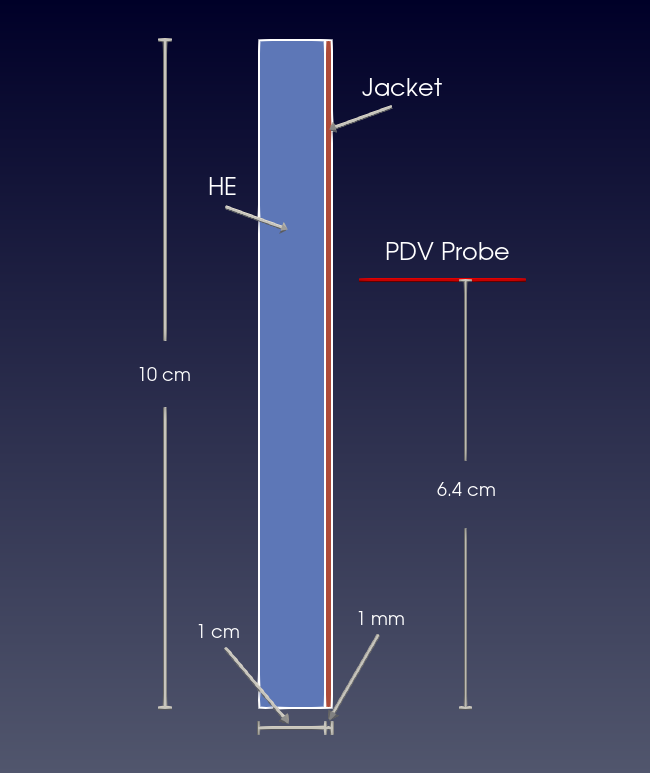}
        \caption{The schematic for the rate stick experiment.} 
    \end{subfigure}
    \begin{subfigure}[t]{0.48\textwidth}
        \centering
        \includegraphics[width=0.9\linewidth]{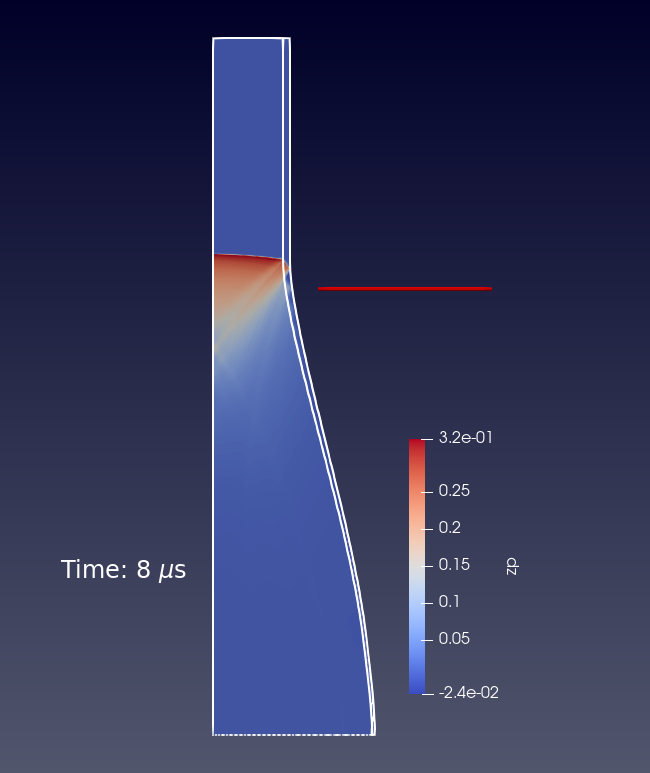}
        \caption{A depiction of the rate stick experiment at the time the PDV velocity reading is taken.}
    \end{subfigure}
    \caption{}
    \label{fig:setup}
\end{figure}

In this section, we describe a series of computer experiments conducted for the PBX 9501 high explosive (HE) and discuss its connection to the present framework. Rate stick experiments, or cylinder tests, are often used to characterize the performance of an HE. These experiments are carried out by detonating a cylindrical stick of HE surrounded by a metal jacket. Photon Doppler Velocimetry (PDV) diagnostics are used to measure the liner velocity at various axial locations to track the progress of the detonation (see \cref{fig:setup}). The collected PDV data can then be used to infer properties of the HE detonation process. In these cases, PDV data far enough away from the detonation point for the detonation front to burn in a steady manner are desirable.

\begin{figure}
    \centering
        \begin{subfigure}[t]{0.31\textwidth}
        \centering
        \includegraphics[width=0.95\linewidth]{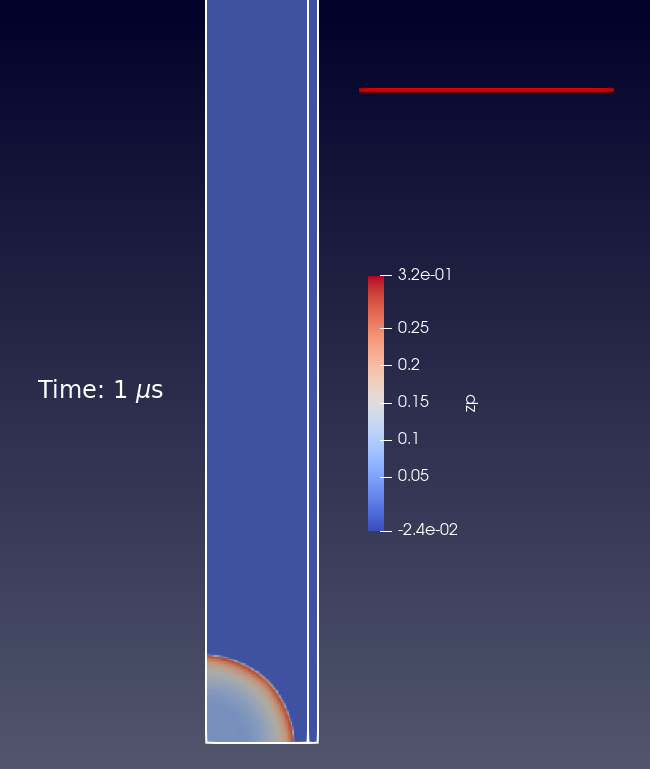}
        \caption{After detonation, a shock wave traverses spherically through the HE.}
    \label{fig:experiment1us}
    \end{subfigure}
    \begin{subfigure}[t]{0.31\textwidth}
        \centering
        \includegraphics[width=0.95\linewidth]{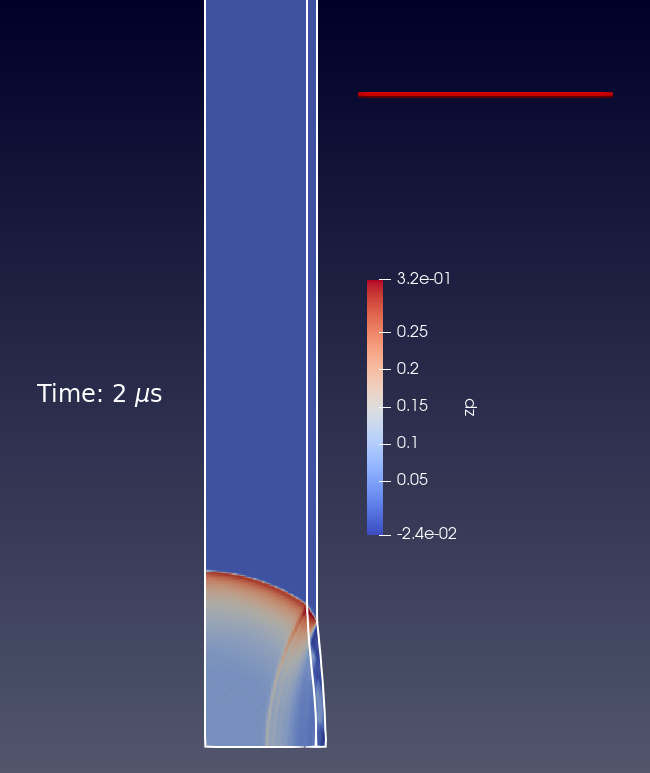}
        \caption{As the shock wave reaches the jacket liner interface, a shock is reflected back towards the detonation product.}
    \label{fig:experiment2us}
    \end{subfigure}
    \begin{subfigure}[t]{0.31\textwidth}
        \centering
        \includegraphics[width=0.95\linewidth]{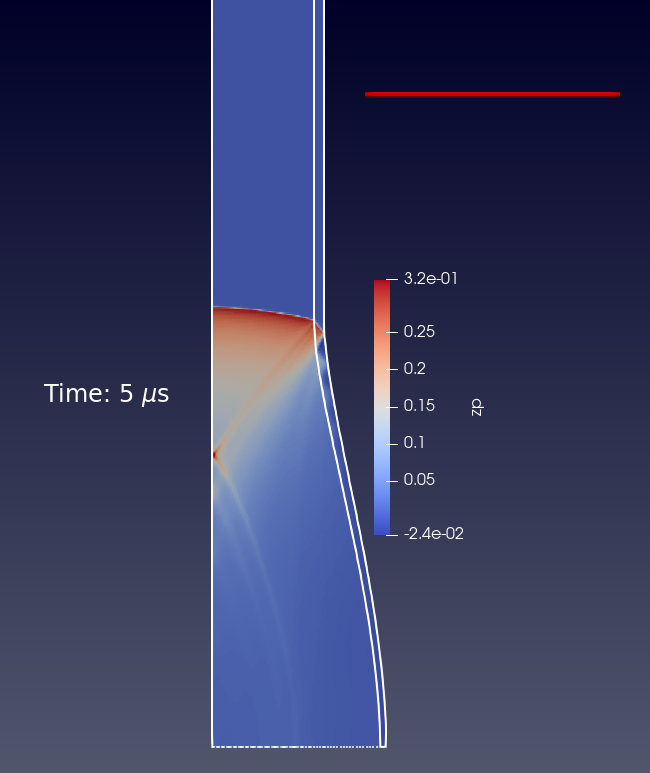}
        \caption{As the shock wave reaches the outside of the jacket liner, the jacket liner begins to expand, and another shock is reflected inwards.}
    \label{fig:experiment5us}
    \end{subfigure}
    \caption{A depiction of the complicated non-linear dynamics governing the rate stick experiment. }
    \label{fig:experiment}
\end{figure}

The rate stick experiment begins with a detonator firing at the center of the bottom face of a cylinder of HE, causing a shock wave to propagate spherically through the HE (\cref{fig:experiment1us}). As the shock passes through the HE, it burns, releasing energy and leaving gaseous detonation products behind.  When the shock reaches the HE-metal interface, part of the shock is transmitted into the metal, and part is reflected back into the detonation products. Once the shock reaches the free boundary of the metal jacket, the jacket begins to expand, and part of the shock is again reflected (\cref{fig:experiment2us}). As these waves traverse through the system, their interactions with interfaces and boundaries yield additional shocks and rarefaction waves, resulting in complex shock structures as shown in \cref{fig:experiment5us}. 
    
A simple two-dimensional schematic of a rate stick experiment and the pressure field at the time of a PDV measurement of interest is shown in \cref{fig:setup}. To simulate these experiments accurately, many different types of physics must be appropriately accounted for. In addition to hydrodynamics, the HE burn, material strength, melt, and damage behavior must be accurately modeled to cover the wide ranges of material responses that may occur in such an experiment. Developed at Los Alamos National Laboratory, the free Lagrangian hydrodynamics code (FLAG) was used to simulate rate stick experiments \citep{crowley2005flag, burton2007lagrangian}. 
    
\begin{table}[!htbp] \centering 
    \caption{Jacket liner materials for the rate stick simulations.}
    \label{tab:jacket_materials}
\begin{tabular}{@{\extracolsep{5pt}} llcc} 
\\[-1.8ex]\hline 
\hline \\[-1.8ex] 
Material & Short Name & Density (g/cc) & Work Hardening \\ 
\hline \\[-1.8ex] 
Aluminum 6061 & al6061 & $2.703$ & $125$ \\ 
Aluminum 7075 & al7075 & $2.804$ & $965$ \\ 
Copper & copper & $8.930$ & $36$ \\ 
Gold & gold & $19.300$ & $49$ \\ 
Gold -- 5\% Copper & gold\_5cu & $18.100$ & $1000$ \\ 
Nickel & nickel & $8.900$ & $46$ \\ 
Stainless Steel 304 & ss304 &  $7.900$ & $43$ \\ 
Stainless Steel 250 & ss250 & $8.129$ & $2$ \\ 
Stainless Steel 4340 & ss4340 &  $7.810$ & $2$ \\ 
Tin & tin & $7.287$ & $2000$ \\ 
Tungsten & tungesten &  $19.300$ & $24$ \\ 
Uranium & uranium & $19.050$ & $2000$ \\ 
Uranium -- 5\% Molybdenum & uranium\_5mo & $18.170$ & $250$ \\ 
Uranium -- 0.75\% Titanium & uranium\_075ti & $18.620$ & $1000$ \\ 
\hline \\[-1.8ex] 
\end{tabular} 
\end{table} 
 
Simulations in FLAG were run for $K=14$ different jacket materials, listed in \cref{tab:jacket_materials}. All liners use the Mie-Gr\"{u}niesen solid equation of state (EoS) \citep{fredenburg2015gruneisen}, the Steinberg-Cochran-Guinan strength model \citep{steinberg1980constitutive}, a modified analytic Lindemann melt model \citep{lawson2009physics}, and a $P_\text{min}$ spall damage model \citep{davison1972continuum}. Nominal parameters for each of the metals were obtained from \cite{steinberg1980constitutive}. 

The FLAG simulation settings for each jacket material constitutes a different physical system with computer model $f_k(\bm x)$, $k=1,\ldots, 14$. We let $\bm x$ denote the parameters of the Jones-Wilkins-Lee (JWL) EoS \citep{lee1968adiabatic} for PBX 9501, which are shared inputs for all computers. The JWL EoS is given by 
\begin{equation}
\label{eq:jwl}
    P(V, e) = A \left( 1 - \dfrac{\omega}{R_1 V} \right) e^{-R_1 V} + B \left( 1 - \dfrac{\omega}{R_2 V} \right) e^{-R_2 V} + \dfrac{\omega \rho_0 e}{V},
\end{equation}
where $\rho_0$, $A, B, R_1, R_2$, and $\omega$ are physical parameters. This EoS is a function of the normalized specific volume $V = \rho_0 / \rho$ and the specific internal energy $e$. \Cref{tab:jwl_pbx9501} gives a range for each input (constructed as nominal value $\pm 3\%$) and the probability distribution for $\bm x$ is taken to be uniform over the hyper-rectangle defined by these bounds. 
\begin{table}[!htbp] \centering 
  \caption{Nominal JWL parameters for PBX 9501 and their uncertainty ranges.} 
  \label{tab:jwl_pbx9501} 
\begin{tabular}{@{\extracolsep{5pt}} ccccc} 
\\[-1.8ex]\hline 
\hline \\[-1.8ex] 
 Input & Parameter & Nominal Value & Lower Bound & Upper Bound \\ 
\hline \\[-1.8ex] 
$x_1$ & $\rho_0$ & $1.840$ & $1.785$ & $1.895$ \\
$x_2$ & $A$ & $8.524$ & $8.268$ & $8.780$ \\ 
$x_3$ & $B$ & $0.182$ & $0.177$ & $0.187$ \\ 
$x_4$ & $R_1$ & $4.550$ & $4.414$ & $4.686$ \\ 
$x_5$ & $R_2$ & $1.300$ & $1.261$ & $1.339$ \\ 
$x_6$ & $\omega$ & $0.380$ & $0.369$ & $0.391$ \\ 
\hline \\[-1.8ex] 
\end{tabular} 
\end{table} 
For each of the $14$ jacket materials, $500$ simulations are performed using a maximin Latin hypercube over the hyper-rectangle defined in \cref{tab:jwl_pbx9501} \citep{park1994optimalLHS, carnell2016package}. In all cases, other HE characteristics such as the detonation velocities are modified to be consistent with the JWL parameterization \citep{weseloh2014jwl}. 

The primary goal is to find subsets of jacket materials which can be viewed as ``neighbors", in the sense that the computer model for neighboring systems behaves similarly when subjected to perturbations in the input space for $\bm x$. In other words, the aim is to find a class (or classes) of computer models for which the gradient is responsive in the same direction of the input space. A secondary goal is to understand the sensitivity of each physical system to the inputs $\bm x$ and to identify differences in sensitivity across systems. 

\subsection{Review of Active Subspaces}

In the active subspace methodology (ASM) of \cite{constantine2015activebook}, the primary goal is to find a projection of the inputs, i.e., $\bm A^\intercal \bm x$ with $\bm A = [\bm a_1 \ \cdots \ \bm a_r]$ and $r \leq p$, such that the output of $f$ is most sensitive to the projected inputs (active variables) $\bm a_i^\intercal \bm x$ ($i=1,\ldots r)$. In particular, the first active direction $\bm a_1 \in \mathbb R^{p\times 1}$ gives the direction which maximizes the expected magnitude of the directional derivative of $f$ with respect to the input distribution $\mu$. Subsequent active directions also maximize this quantity subject to the constraint that they are orthogonal to each of the previous active directions. This is accomplished by first finding the expected outer product of the gradient of $f$, 
\begin{equation}
    \label{eq:Cmatrix}
    C_f = \mathbb E\left[\nabla f(\bm x) \nabla f(\bm x)^\intercal\right] = \int \nabla f(\bm x) \nabla f(\bm x)^\intercal d\mu.
\end{equation}
The matrix $\bm C_f$ provides a wealth of valuable information, and its eigenvalue decomposition $\bm C_f = \bm W\bm\Lambda\bm W^\intercal$ can be revealing. For some threshold $\tau > 0$, let $r$ be the number of eigenvalues ($\lambda_i = \bm\Lambda_{ii}$) which are at least $\tau$ and let $\bm A \in \mathbb R^{p\times r}$ be the first $r$ columns of $\bm W$. The column space of $\bm A$ is called the active subspace, the $i^{th}$ column of $\bm A$ (denoted $\bm a_i$) is the $i^{th}$ active direction and the linear transformation $\bm a_i^\intercal \bm x$ is the $i^{th}$ projected input. Global sensitivity metrics for each native input $(i=1,\ldots,p)$ can be defined as
\begin{equation}
    \label{eq:activityscores}\alpha_i(q) = \sum_{j=1}^q\lambda_j w_{i,j}^2, 
\end{equation}
where $w_{i,j}$ is the $(ij)^{th}$ entry of $\bm W$ and $q \leq p$ \citep{constantine2017global}. The special case $q = 1$ is of notable importance and is often referred to as the {\it activity score} for input $x_i$.

\section{Co-Active Subspace Methods}
\label{sec:CASM}
$K$ pairwise adjacent functions, $f_1, \ldots, f_K$, are of interest
\begin{equation}
    f_k(\bm x): \mathcal X \mapsto \mathbb R, k=1,\ldots, K,
\end{equation}
with $\mathcal X \subset \mathbb R^p$. The input vector $\bm x$ is equipped with a probability distribution $\bm x \sim \mu$, and therefore the gradient $\nabla f_k(\bm x)$ is a random p-vector because of its dependence on $\bm x$. We assume that each function is deterministic, differentiable, and Lipschitz continuous and we define the expected outer product
\begin{equation}
\label{eq:Cij}
    \bm C_{k\ell} = \mathbb E_\mu\left[ \nabla f_k(\bm x) \nabla f_\ell(\bm x)^\intercal \right] = \int \nabla f_k(\bm x) \nabla f_\ell(\bm x)^\intercal d\mu,
\end{equation}
and the expected inner product as 
\begin{equation}
    \label{eq:tij}
    t_{k\ell} = \text{trace}(\bm C_{k\ell}) = \mathbb E_\mu\left[ \nabla f_k(\bm x)^\intercal \nabla f_{\ell}(\bm x) \right].
\end{equation}
This generalizes the univariate ASM matrix from \cref{eq:Cmatrix}, in the sense that $\bm C_{kk} = \bm C_{f_k}$. For notational brevity, we write $\bm C_{kk} \stackrel{\triangle}{=} \bm C_k$ and $t_{kk} \stackrel{\triangle}{=} t_k$.  When $f_k \neq f_\ell$, we note that $\bm C_{k\ell}$ is not generally symmetric. It can be useful to define the symmetrized matrix as
\begin{equation}
    \label{eq:Vij}
    \bm V_{k\ell} = \frac{\bm C_{k\ell} + \bm C_{\ell k}}{2}
\end{equation}
and to note that for any unit vector $\bm w$ the expected product of derivatives in the direction $\bm w$ is 
\begin{equation} \label{eq:CVcompare}
    \bm w^\intercal C_{k\ell} \bm w = 
    \bm w^\intercal C_{\ell k} \bm w = 
    \bm w^\intercal V_{k\ell} \bm w. 
\end{equation}
Symmetrizing is useful because it allows us to study the eigendecomposition of $\bm V_{k\ell}$, drawing many direct comparisons to traditional active subspace methods. This approach, however, masks which function depends on which variables. In order to qualitatively recover this information, the matrices $\bm V_{k\ell}$, $\bm C_k$, and $\bm C_\ell$ must be studied simultaneously. An example of this is given in \cref{sec:cosenspbx} and \cref{fig:coactivity1} and \cref{fig:coactivity2}, where the ``co-activity" of an input with respect to two functions should be viewed relative to the ``activity" of the input for each function individually. 

Finally, we note that \cite{lee2019modified} recommends a modification to the traditional active subspace, showing that it can improve the discovered subspace when the underlying function is quadratic. Although we do not explore this fully here, this modification can also be applied to the co-active subspace method by replacing $\bm C_{k\ell}$ with
\begin{equation}
\label{eq:modifiedasm}
    \tilde{\bm C}_{k\ell} = \mathbb E_\mu\left[ \nabla f_k(\bm x) \nabla f_\ell(\bm x)^\intercal \right] + \mathbb E_\mu\left[\nabla f_k(\bm x) \right]\mathbb E_\mu\left[\nabla f_\ell(\bm x) \right]^\intercal.
\end{equation}

\subsection{Similarity Under Perturbation: Concordance and Discordance}
\label{sec:concordance}
If two functions respond similarly to directional perturbations in the input space, then they are close to each other in some abstract, but important sense. Studying changes in the output of the functions is useful (and should be done), but it is not sufficient by itself to capture all meaningful similarities and fully distinguish between all discrepancies. For instance, the measure $\int |f_1(\bm x) - f_2(\bm x)|^2 d\mu$ can fail to recognize the close relationship between $f_1(\bm x)$ and  $f_2(\bm x) = a + bf_1(\bm x)$. Looking at the gradients of $f_k$ can provide useful information, so the analysis often benefits from supplementing with measures based on active directions. The Grassmann distance \citep{ye2016schubert} is a useful tool for finding the distance between subspaces, but this approach by itself can fail to solve the problem of interest in some very simple settings. As an extreme example, the Grassman distance is exactly zero when $f_2(\bm x) = -f_1(\bm x)$. It can also understate how poorly two functions track at the same $\bm x$ when $f_2(\bm x) = f_1(\bm x + \bm \epsilon)$ for fixed $\bm \epsilon \in {\mathbb R}^p$. For example, the functions $f_1(\bm x) = \sin(c\bm x^\intercal \bm 1)$ and $f_2(\bm x) = \sin(c(\bm x+\bm\epsilon)^\intercal \bm 1)$ will have the similar active subspaces for large values of $c$ and thus the Grassman distance will be small. Yet, the behavior of these functions under perturbation can differ considerably at different locations in the input space. Instead, we propose that, when comparing two non-constant functions, the analysis should be supplemented by a single number, referred to as the {\it concordance} of $f_k$ and $f_\ell$, 
\begin{equation}
\label{eq:conc}
    \kappa_{k\ell} = \text{conc}(f_k, f_\ell) = \frac{t_{k\ell}}{\sqrt{t_kt_\ell}}, \quad f_k, f_\ell \not\propto 1.
\end{equation}
Since by Cauchy-Schwarz, $t_{k\ell}^2 \leq t_kt_\ell$, it invariably holds that $\text{conc}(f_k, f_\ell) \in [-1, 1]$. There is a clear analogy here to Pearson's correlation, as the concordance is $1$ if and only if $f_\ell(\bm x) = a + |b|f_k(\bm x)$ and it is equal to $-1$ if and only if $f_\ell(\bm x) = a - |b|f_k(\bm x)$ (where the equalities hold almost everywhere with respect to $\mu$). Likewise, the concordance is exactly $0$ when the gradient of $f_k$ is almost everywhere orthogonal to the gradient of $f_\ell$. It is often convenient to have a formal distance metric, as we will see in \cref{sec:HE}, so we define {\it discordance} as 
\begin{equation}
\label{eq:discord}
    \text{discord}(f_k, f_\ell) = \sqrt{\frac{1-\text{conc}(f_k, f_\ell)}{2}},
\end{equation}
which satisfies the necessary properties of a pseudo-metric (see Section 3 of the SM). Note that if concordance is the only metric of interest, computation time can be saved by calculating only the diagonal of the $\bm C_{k\ell}$ matrix. 

Concordance is appealing as a highly compact summary of the alignment of the gradient spaces of two functions, and the restriction of concordance to the interval $[-1, 1]$ is convenient for interpretation. However, deciding what value of concordance corresponds to ``close alignment" of two models is highly context dependent. For example, in most of our applications of interest (multi-physics simulations), it is usually the case that all adjacent models pairs are relatively concordant so that even a small departure from $1$ may be meaningful. Much smaller or even negative values of concordance may be routine in other applications.

\subsubsection{Simple polynomial example}
\begin{figure}[t]
    \caption{}
    \label{fig:simple_conc}
    \centering
    \begin{subfigure}[t]{0.48\textwidth}
    \centering
    \includegraphics[width=0.98\linewidth]{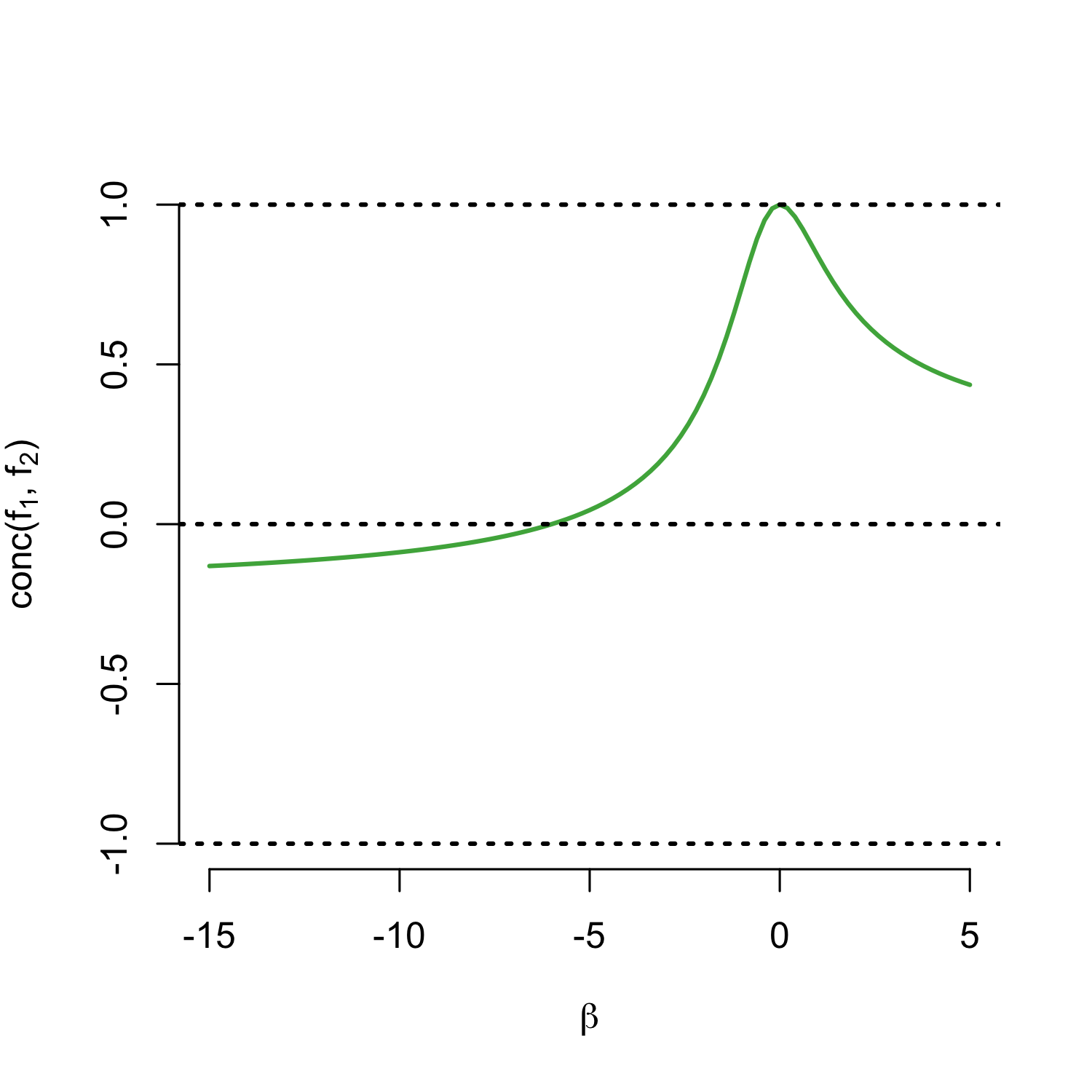}
    \caption{Concordance between $f_1$ and $f_2$ in the simple polynomial example, as a function of $\beta$. }
    \label{subfig:conc}
    \end{subfigure}
    \begin{subfigure}[t]{0.48\textwidth}
    \centering
    \includegraphics[width=0.98\linewidth]{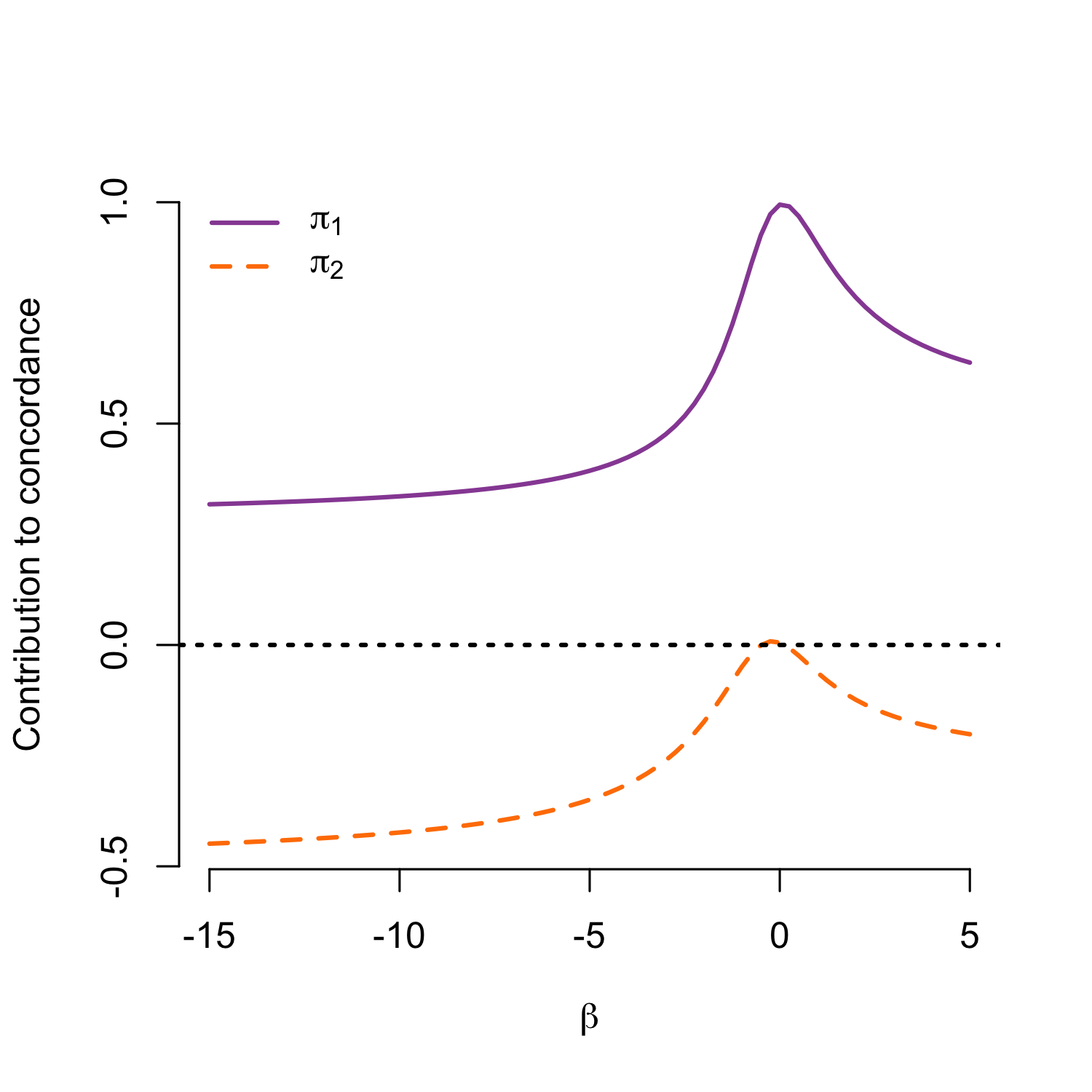}
    \caption{The concordance is the sum of the contributions of the first and second co-active directions, $\pi_1 + \pi_2$.}
    \label{subfig:contribution}
    \end{subfigure}
\end{figure}

Code to reproduce all figures and quantities in this section is provided at \url{https://github.com/knrumsey/concordance/inst/CoASM/simple_poly_concordance.R}. For illustration throughout this section, consider the functions
\begin{equation}
\label{eq:simple_example}
\begin{aligned}
f_1(\bm x) &= x_1^2 + x_1 x_2 \\
f_2(\bm x) &= x_1^2 + x_1 x_2 + \beta x_2^3, \ \beta \in \mathbb R, 
\end{aligned}
\end{equation}
and assume that $x_1, x_2 \stackrel{\text{iid}}{\sim} \text{Unif}(0, 1)$. Here, we have 
\begin{equation}
\label{eq:simple_results1}
\begin{aligned}
    \bm C_1 &= \frac{1}{180}\begin{bmatrix}
    480 & 165 \\
    165 & 60
    \end{bmatrix}, \quad\quad\quad
    &&\bm C_2 = \frac{1}{180}\begin{bmatrix}
    480 & 165 + 315\beta \\
    165 + 315\beta & 60 + \beta(324\beta+180)
    \end{bmatrix}, \\[1.2ex]
    \bm C_{12} &= \frac{1}{180}\begin{bmatrix}
    480 & 165 + 315\beta \\
    165 & 60 + 90\beta
    \end{bmatrix}, \quad\quad\quad 
    &&\bm V_{12} = \frac{1}{180}\begin{bmatrix}
    480 & 165 + \frac{315}{2}\beta \\
    165+ \frac{315}{2}\beta & 60 + 90\beta
    \end{bmatrix}.
\end{aligned}
\end{equation}
Note that when $\beta = 0$, we have $f_1 = f_2$ and all matrices in \cref{eq:simple_results1} become identical. Now, we write
\begin{equation}
    \begin{aligned}
    t_1 = 3, \quad\quad
    t_2 &= 3 + \beta\left(\frac{9}{5}\beta + 1\right) \quad\quad
    t_{12} = 3 + \frac{1}{2}\beta, \\[1.2ex]
    \text{conc}(f_1, f_2) &= \frac{3+\frac{\beta}{2}}{\sqrt{9 + 3\beta\left(\frac{9}{5}\beta + 1\right)}}.
    \end{aligned}
\end{equation}
The concordance of $f_1$ and $f_2$, as a function of $\beta$, is shown in \cref{subfig:conc}. When $\beta$ is relatively small (e.g., $\beta = 1/2$) the functions are intuitively very similar and the concordance is high. As $\beta$ gets large in magnitude however, the functions become much less concordant. For example, when $\beta = 1/2$ the concordance is $0.944$, but the concordance is just $0.551$ for $\beta=3$ and $-0.131$ for $\beta = -12$. 

\subsection{Co-Active Directions}
As with traditional ASMs, there is much to learn from the eigenvalue decomposition of the expected outer product. To take advantage of the useful properties of the eigendecomposition on symmetric matrices, we focus on the symmetrized matrix $\bm V_{k\ell}$ rather than $\bm C_{k\ell}$. Note that, unlike in the one-function case, $\bm V_{k\ell}$ is potentially indefinite and thus $\bm V_{k\ell}$ could have negative eigenvalues. We define
\begin{equation}\label{eq:Vdecomp}
    \bm V_{k\ell} = \bm W_{k\ell}\bm \Lambda_{k\ell} \bm W_{k\ell}^\intercal,
\end{equation}
where $\bm W_{k\ell}$ is a $p\times p$ orthonormal matrix of eigenvectors $\bm w_i^{(k\ell)} (i=1,\ldots, p)$ and $\bm\Lambda_{k\ell}$ is a diagonal matrix of eigenvalues $\lambda_i^{(k\ell)} = \left(\bm\Lambda_{k\ell}\right)_{ii}$ ordered from largest ($\lambda_1^{(k\ell)}$) to smallest ($\lambda_p^{(k\ell)}$).

We refer to the leading columns of $\bm W_{k\ell}$ as the {\it co-active directions}, noting that (by \cref{eq:CVcompare} and properties of an eigendecomposition) the first such direction, $\bm w_1^{(k\ell)}$, maximizes the expected product of directional derivatives, and the subsequent directions maximize the same quantity while maintaining orthogonality between all previous co-active directions. That is, 
\begin{equation}
\label{eq:coactivedir}
\begin{aligned}
    \bm w_1^{(k\ell)} &= \argmax_{\bm w: \|\bm w\| = 1} \int \left(\bm w^\intercal \nabla f_k(\bm x)\right)\left(\bm w^\intercal \nabla f_\ell(\bm x)\right) d\mu \\[1.2ex]
    \bm w_j^{(k\ell)} &= \argmax_{\stackrel{{\bm w: \|\bm w\| = 1}}{\bm w^\intercal \bm w_i^{(k\ell)}=0, (i < j)}} \int \left(\bm w^\intercal \nabla f_k(\bm x)\right)\left(\bm w^\intercal \nabla f_\ell(\bm x)\right) d\mu, \quad j=2,\ldots,p.
\end{aligned}
\end{equation}
Moreover, the eigenvalues satisfy
\begin{equation}
    \lambda_i^{(k\ell)} = \bm w_i^{(k\ell)\intercal} \bm V_{k\ell} \bm w_i^{(k\ell)} = \int \left(\bm w_i^\intercal \nabla f_k(\bm x)\right)\left(\bm w_i^\intercal \nabla f_\ell(\bm x)\right) d\mu, 
\end{equation}
which indicates that $\lambda_i$ is large (in magnitude) if the expected value of the product of directional derivatives is large (in magnitude). That is, the $i^{th}$ eigenvalue is large only if both functions are active in the direction of the $i^{th}$ eigenvector in some region of the input space. Positive eigenvalues indicate that upon perturbation in the direction of $\bm w_i^{(k\ell)}$, both functions change in the same way (i.e., both increase, or both decrease) and negative values imply the opposite (i.e., one function increases while the other decreases). We can tie this back to concordance by noting that, since $t_{k\ell} = tr(\bm V_{k\ell}) = tr(\bm\Lambda_{k\ell})$ (and because $t_k, t_\ell \geq 0$), the concordance of $f_k$ and $f_\ell$ can be negative (or positive) only when the sum of the eigenvalues is negative (or positive). Eigenvalues which are small in magnitude can occur in one of two ways. Most commonly, this occurs because (at least) one of the two functions $f_k$ or $f_\ell$ is constant or nearly constant in the direction of the corresponding eigenvector, for every point in the input space. This can also occur if the effect of moving along the active direction has opposite effect on the response for various regions of the input space. 

To aid in the interpretation of the eigenvalues, it is sometimes helpful to view them in their standardized form
\begin{equation}
\label{eq:contrib}
\pi_i^{(k\ell)} = \frac{\lambda_i^{(k\ell)}}{\sqrt{t_kt_\ell}}.
\end{equation}
We call these quantities {\it contributions} because they indicate the degree to which variation along each co-active direction contributed to the concordance. In particular, note that $\sum_{i=1}^p\pi_i^{(k\ell)} = \text{conc}(f_k, f_\ell)$. In the case where $f_k = f_\ell$, the contributions are certain to be positive and so $\pi_i^{(k)}$ reduces to the proportion of the total gradient. In the case where $f_k \neq f_\ell$, the contributions cannot be interpreted as proportions, but their signs and magnitudes still carry valuable and interpretable information about the co-active directions.

\subsubsection{Low Dimensional Representation of Adjacent Models}
Much of the literature for ASM motivates active directions through a Gaussian Poincar\'e inequality that bounds the mean squared error (MSE) in approximating a target function by its conditional expectation on a linear subspace of the inputs.  Active directions minimize the Poincar\'e bound. For example, if $\mu = {\cal N}_p(\bm m, \bm \Sigma)$ then Proposition~2.5 of \cite{zahm2020gradient} bounds the approximation MSE on the $r$ dimensional subspace ${\cal B} = \text{span}(\bm B) \subset {\mathbb R}^p$ where the $r$ columns of $\bm B$ are linearly independent input directions:
\begin{align}
    \label{eq:approximtionBound}
    \int (f(\bm x) - \hat f(\bm x))^2 d\mu
    & \leq
    \text{trace}\left( \bm \Sigma (\bm I - \bm P_B) \bm C_f (\bm I - \bm P_B) \right) ,
\end{align}
where $\bm C_f$ was defined in \cref{eq:Cmatrix}, 
$\bm P_B=\bm B (\bm B^\intercal \bm B)^{-1} \bm B^\intercal$ 
is the projection matrix onto $\cal B$, and 
$\hat f(\bm x) = 
   {\mathbb E}_{\bm X\sim\mu}[f(\bm X) 
   \, | \, 
   \bm P_B \bm X = \bm P_B \bm x ] $ 
is the conditional expectation on $\cal B$.   
Transforming to canonical coordinates $\tilde{\bm x} = \bm \Sigma^{-1/2} \bm x$ and setting $\bm B$ to be the first $r$ resulting AS directions minimizes the bound for a linear subspace of dimension $r$ 
\citep[Proposition~2.6, simplified]{zahm2020gradient}.
On the other hand, setting $\bm B$ to the leading columns of $\bm W_{k\ell}$ (i.e. \cref{eq:Vdecomp}) produces approximation bounds for $f=f_k$ and $f=f_\ell$ on the co-active subspace and, in general, these bounds will be larger than the optimal AS bounds for the same subspace dimension. 

But co-active directions have a different goal; they capture the subspace on which the gradients of two functions are best aligned.  In a high concordance situation, the co-active subspace will also be suitable for approximating the two functions, but this is not the primary purpose. Along these lines, Section~\ref{sec:lowD_JWL} compares approximation errors on a variety of different subspaces in the high explosive application.  

To determine an appropriate size of the lower dimensional subspace, one can adopt the common approach for single-function active subspaces, which is to look for gaps in the eigenvalues. This can be done visually or via a sequential testing procedure \citep{ma2013review}. The only distinction for co-ASM is the possibility of negative eigenvalues, and so they should be ordered by their absolute magnitude rather than the conventional ordering. 

\subsubsection{Simple polynomial example}
In the high-concordance case of $\beta = 1/2$, the first active directions for $f_1$ and $f_2$ are, respectively, $\bm w_1^{(1)} \approx [0.945 \ 0.327]^\intercal$ and $\bm w_1^{(2)} \approx [0.825 \ 0.566]^\intercal$. These active directions are similar, but it does correctly indicate that $x_2$ is relatively more important for $f_2$ than for $f_1$. The first co-active direction is given by $\bm w_1^{(12)} \approx [0.907 \ 0.422]^\intercal$, the optimal compromise between the active directions. The contribution vector is $\bm\pi^{(12)} = [0.9518 \ \ -0.0077]$. From these quantities, we learn that moving in the direction $0.907 x_1 + 0.422x_2$ leads to substantial changes on average in magnitude for both $f_1$ and $f_2$ (in the same direction). Moving orthogonal to this direction leads to less substantial, but non-zero, changes in the output of $f_1$ and $f_2$ in opposite directions. The contributions, $\pi_1^{(12)}$ and $\pi_2^{(12)}$, are plotted in \cref{subfig:contribution} as a function of $\beta$. 

\subsection{Co-sensitivity Analysis}

We now turn our attention to assessing the sensitivity of the functions to the native inputs. By straightforward extension of the activity score sensitivity metric given in \cref{eq:activityscores}, we define the {\it signed co-activity score} for input $x_i$ with respect to $f_k$ and $f_\ell$ to be 
\begin{equation}
\label{eq:signed}
    \alpha_i^{(k\ell)}(q) = \sum_{j=1}^q\lambda^{(k\ell)}_{i_j} \left(w^{(k\ell)}_{i,i_j}\right)^2,
\end{equation} 
where the indices $i_1, \ldots i_p$ represent a permutation of the set $\{1, \ldots, p\}$ such that $$\left |\lambda_{i_1}^{(k\ell)}\right | \geq \left |\lambda_{i_2}^{(k\ell)}\right | \geq \ldots \geq \left |\lambda_{i_p}^{(k\ell)}\right |.$$ This extra step is necessary since the magnitude of the eigenvalues signals the importance of an active direction, regardless of sign. Following the advice of \cite{constantine2017global} we recommend setting $q=1$, but it is also sometimes reasonable to set $q = r$ where $r$ is the number of eigenvalues which exceed some threshold $\tau > 0$ in absolute value. The signed co-activity score can be positive or negative, and this single bit of information carries useful information about the relationship of the corresponding input to the functions $f_k$ and $f_\ell$. When the $i^{th}$ co-activity score is negative, the net effect of input $x_i$ is positive for one of the functions and negative for the other. When $q > 1$, it is possible for the terms in \cref{eq:signed} to cancel out, masking the importance of a particular variable. To account for this, we introduce the {\it unsigned co-activity score} for input $x_i$ with respect to $f_k$ and $f_\ell$ to be 
\begin{equation}
\label{eq:unsigned}
    \tilde\alpha_i^{(k\ell)} = \sum_{j=1}^q\left |\lambda^{(k\ell)}_{i_j}\right | \left(w^{(k\ell)}_{i,{i_j}}\right)^2.
\end{equation}
This sensitivity index can be interpreted like the traditional activity score, in the sense that large positive values indicate that a variable is highly active in functions $f_k$ and $f_\ell$, although the ``direction" of the activity is lost. When $q=1$, the signed and unsigned co-activity scores can differ only in their sign, e.g., $\alpha_i^{(k\ell)} = \pm \tilde\alpha_i^{(k\ell)}$ -- a fact that fails to hold true for $q > 1$.

\begin{figure}[t]
    \centering
    \begin{subfigure}[t]{0.31\textwidth}
    \label{fig:coacts}
    \centering
    \includegraphics[width=0.98\linewidth]{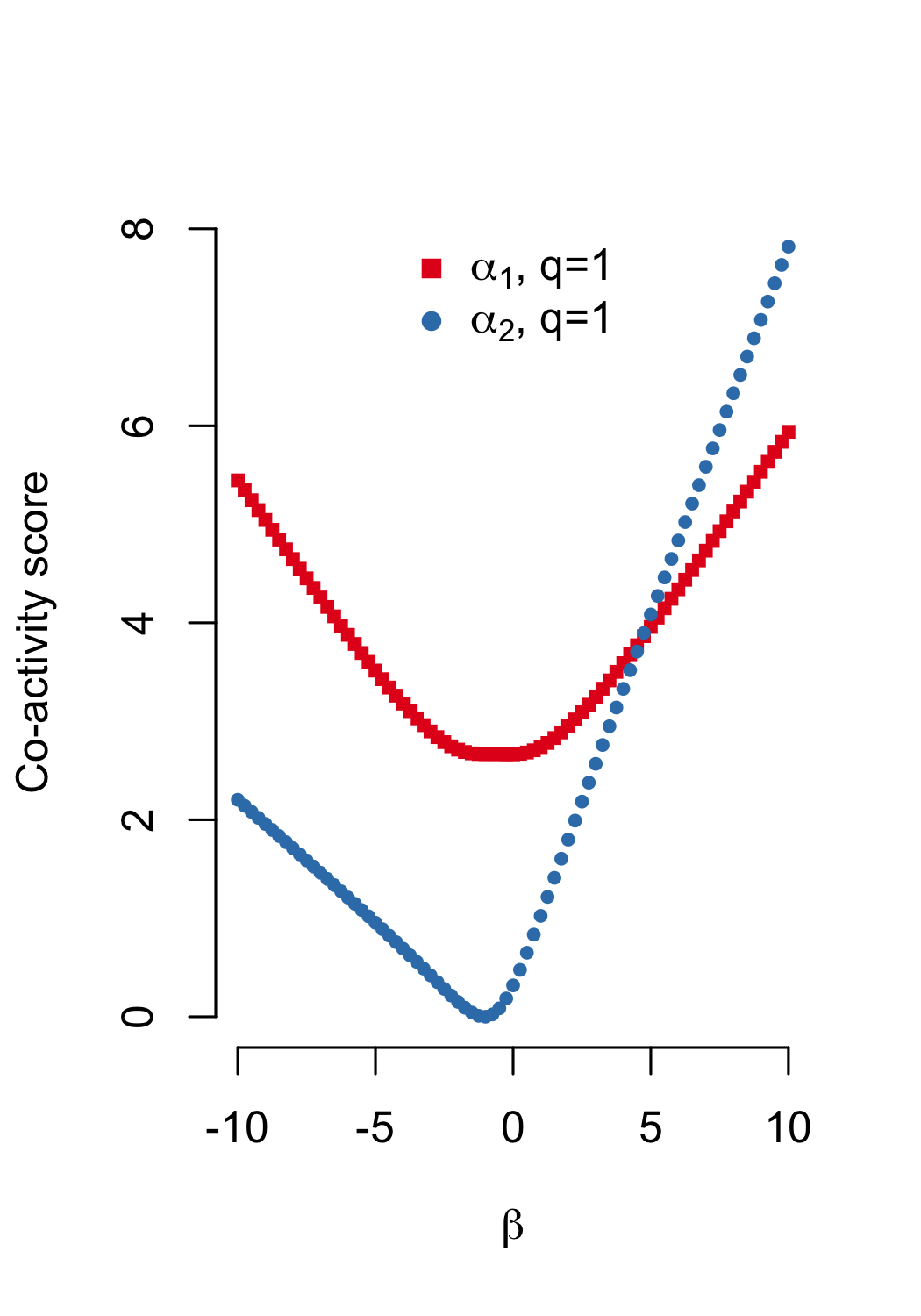}
    \caption{Co-activity scores for the native inputs $x_1$ and $x_2$ as a function of $\beta$. When $q=1$, the signed and unsigned scores are equivalent (up to sign).}
    \end{subfigure}
    \begin{subfigure}[t]{0.31\textwidth}
    \label{fig:coactu}
    \centering
    \includegraphics[width=0.98\linewidth]{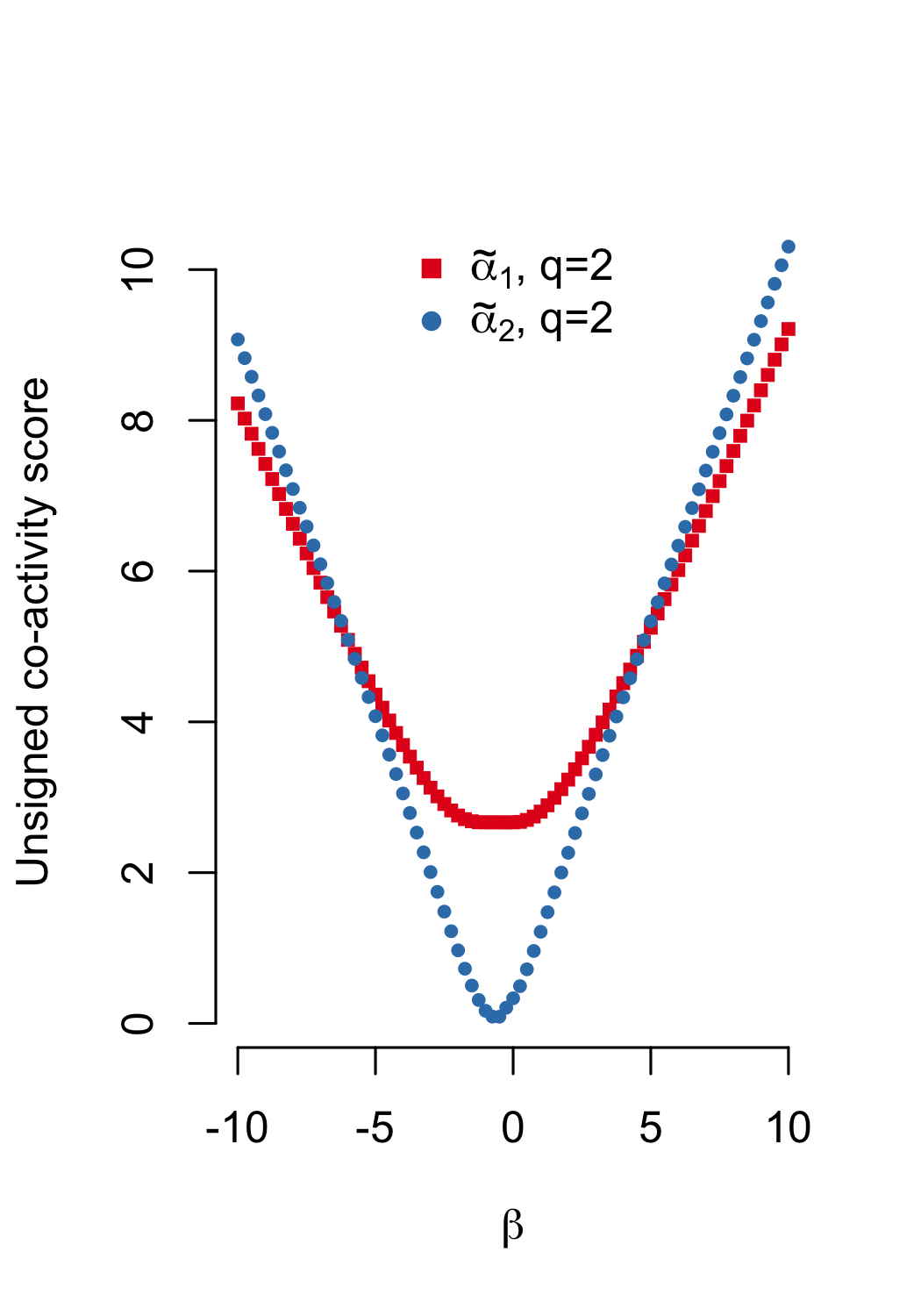}
    \caption{Unsigned co-activity scores (with $q=2$) for the native inputs $x_1$ and $x_2$ as a function of $\beta$. }
    \end{subfigure}
    \begin{subfigure}[t]{0.31\textwidth}
    \label{fig:coactabs}
    \centering
    \includegraphics[width=0.98\linewidth]{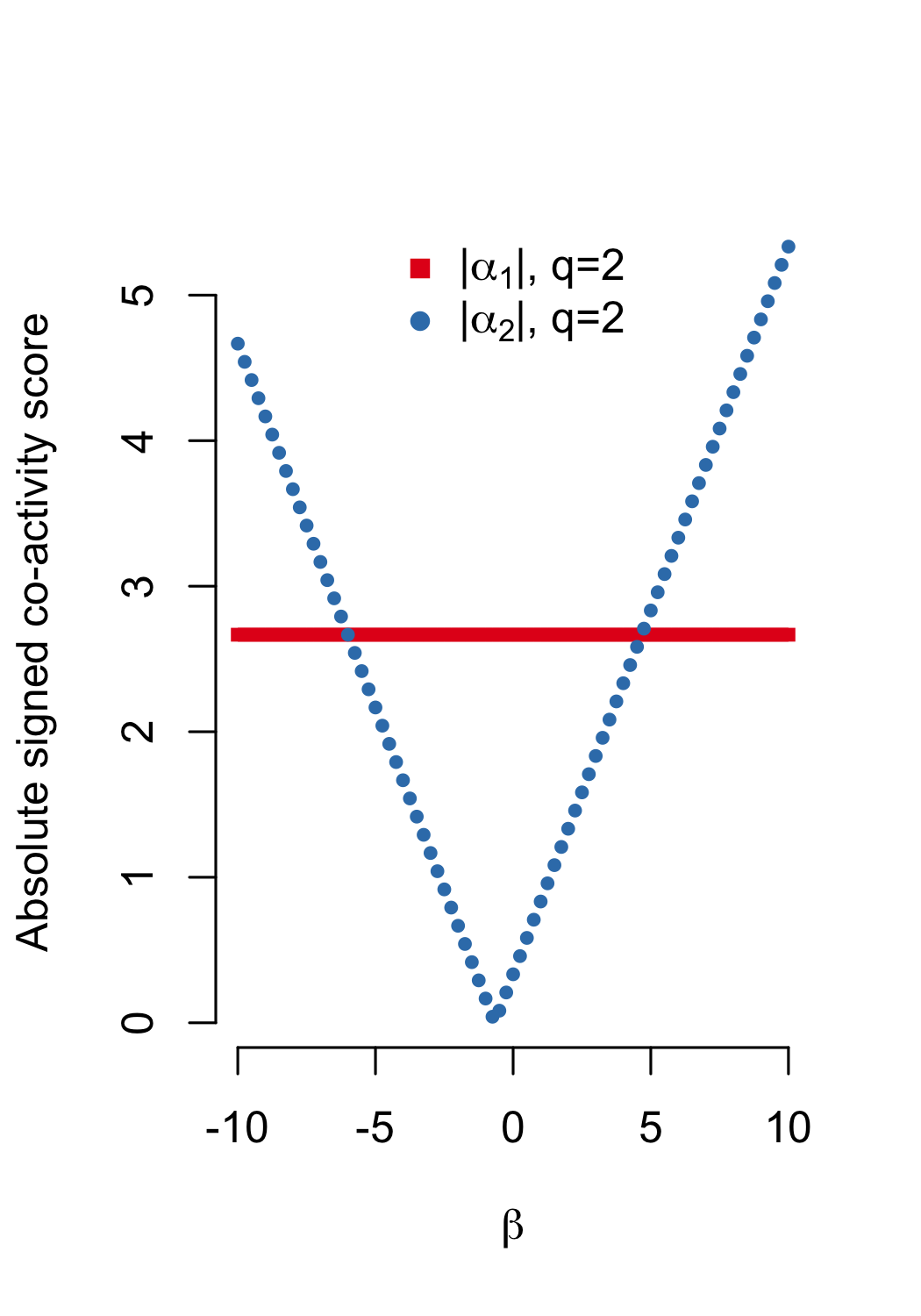}
    \caption{Absolute value of the signed co-activity scores (with $q=2$) for the native inputs $x_1$ and $x_2$ as a function of $\beta$. }
    \end{subfigure}
    \caption{}
    \label{fig:activity}
\end{figure}

\subsubsection{Simple polynomial example}
Some co-activity scores are shown for the simple polynomial example in \cref{fig:activity}. When $q=1$, the absolute-signed and unsigned co-activity scores are equivalent, and they demonstrate that $x_1$ is jointly more important to the functions for $\beta \stackrel{\sim}{\leq} 5$ and $x_2$ is relatively more active otherwise. In the $q=2$ case, the unsigned activity scores are plotted in \cref{fig:coactu} and the absolute value of the signed activity scores are plotted in \cref{fig:coactabs}. According to both of these metrics, $x_2$ becomes more co-active for $f_1$ and $f_2$ when $|\beta| \stackrel{\sim}{\geq} 5$. 

\subsection{Efficient Numerical Approximation}
\label{sec:implementation}
So far, we have been working with an extremely simple example which has allowed for closed form analysis of the co-active subspace, but most interesting and important problems will not be so tractable. Thus, numerical tools are needed to perform these analyses accurately and efficiently. In particular, we need a way to estimate the matrix $\bm C_{k\ell}$ when it is not available analytically, as is usually the case. This is the same computational and theoretical bottleneck faced in traditional active subspace methodology, where the majority of the literature has relied on Monte Carlo (MC) estimation \citep{constantine2014computing, constantine2015activebook}. In the MC approach, we assume that the gradient of each $f_k$, 
$$\nabla f_k(\bm x): \mathbb R^p \mapsto \mathbb R^p,$$
can be calculated (or estimated) quickly and accurately. Then, we simply iterate between (i) sampling $\bm x_b \sim \mu$ and (ii) calculating $\hat {\bm C}^{(k\ell)}_b = \nabla f_k(\bm x_b) \nabla f_\ell(\bm x_b)^\intercal$ for $b = 1, \ldots, B$; the final estimator can now be constructed as
\begin{equation}
\label{eq:MC_C}
\hat{\bm C}_{k\ell} = \frac{1}{B}\sum_{b=1}^B\hat {\bm C}^{(k\ell)}_b.
\end{equation}
Although this approach is appealing for its simplicity, it is associated with many notable drawbacks. For high-dimensional functions (e.g., large $p$) the method can be slow to converge and does not easily lend itself to uncertainty estimation. Further, it requires computer experiments to be tailored specifically to the probability distribution $\mu$, which disallows the use of previously existing data sets and does not permit meta-analysis over multiple distributions (as in robust Bayes; \cite{berger1990robust}).  Importance weighting could be incorporated into the procedure for this purpose, but this is seldom an efficient use of computer resources.

A recent alternative \citep{wycoff2021sequential, rumsey2023discovering}, is based on fitting an emulator (or surrogate-model) to data in order to learn the input-output map defined by each $f_k$.  Conditional on this statistical emulator,the $\bm C_{k\ell}$ matrix can be efficiently and accurately computed. In this paper, we generalize the approach discussed by \cite{rumsey2023discovering}, which has been shown to perform well in high-dimensional (large $p$) settings. The first step is to represent each function $f_k, (k=1,\ldots, K)$ as closely as possible by a model of the form,
\begin{equation}
\label{eq:mars}
    f_k(\bm x) \approx \gamma_{k0} + \sum_{m=1}^{M_k} \gamma_{km}\prod_{i=1}^p\left[s_{kim}(x_i-t_{kim})\right ]_+^{u_{kim}},
\end{equation}
where $[x]_+ = \text{max}(x, 0)$ is called a hinge function, $s_{kim} \in \{-1, 1\}$ is a sign, $t_{kim}$ is a knot and $u_{kim} \in \{0, 1\}$ is a variable inclusion indicator. This linear combination of multivariate splines is known to have the universal representation property \citep{lin1992canonical}.  A popular algorithm for fitting models of the form in \cref{eq:mars} is known as multivariate adaptive regression splines (MARS) \citep{friedman1991, denison1998bayesian}. The Bayesian additive spline surfaces (BASS) algorithm of \cite{francom2020bass} represents a modernized version of the algorithm (with parallel tempering, $g$-priors, flexible likelihoods, and cleverly constructed proposal distributions) which has been successfully applied to a wide range of UQ problems \citep{francom2018sensitivity, francom2019calibration, rumsey2024generalized, nott2005} and has proven to be a highly competitive emulator in recent comparisons \citep{collins2022bayesian, hutchings2023comparing, rumsey2023localized}. For the sake of brevity, we leave additional details to the references, especially \cite{francom2018sensitivity} and \cite{francom2020bass}. 

We note that a similar approach can be developed for alternative emulators (including Gaussian processes and additive regression trees), but BASS has been shown to be effective and efficient; see \cite{rumsey2023discovering} for a detailed discussion and comparison of alternatives for a single-function active subspace.

Given two functions $f_k$ and $f_\ell$, we can write the $(ij)^{th}$ element of $\bm C_{k\ell}$ as 
\begin{equation}
    \bm C_{k\ell}[i,j] = \int \frac{\partial f_k}{\partial x_i}(\bm x)\frac{\partial f_\ell}{\partial x_j}(\bm x) \mu(\bm x) d\bm x
\end{equation}
This high-dimensional integral will be intractable for most problems of interest, but by representing each function in the form of \cref{eq:mars}, we can generalize the calculations of \cite{rumsey2023discovering} and write 
\begin{equation}
\label{eq:Capprox}
    \bm C_{k\ell}[i,j] = 
    \begin{dcases}
    \sum_{m_1=1}^{M_k}\sum_{m_2=1}^{M_\ell}\gamma_{km_1}\gamma_{\ell m_2}I_{3,k\ell}^{(i)}[m_1, m_2]\prod_{i^\prime \neq i} I_{2,k\ell}^{(i^\prime)}[m_1, m_2], & i = j \\
    \sum_{m_1=1}^{M_k}\sum_{m_2=1}^{M_\ell}\gamma_{km_1}\gamma_{\ell m_2} I_{1,k\ell}^{(i)}[m_1, m_2]I_{1,\ell k}^{(j)}[m_2, m_1]\prod_{i^\prime \not\in\{i,j\}} I_{2,k\ell}^{(i^\prime)}[m_1, m_2], & i \neq j,
    \end{dcases}
\end{equation}
Where $I_{1,k\ell}^{(i)}[m_1, m_2]$, $I_{2,k\ell}^{(i)}[m_1, m_2]$, and $I_{3,k\ell}^{(i)}[m_1, m_2]$ are univariate integrals which can be solved in closed form, so long as $\mu(\bm x) = \prod_{i=1}^p\mu_i(x_i)$, corresponding to statistical independence of the $x_i$. To construct $\bm C_{k\ell}$ in its entirety, we need to evaluate $4pM_kM_\ell$ such integrals, but the closed form availability of the solution makes the algorithm incredibly fast in practice. Additional details, including the tractable solution to $I_{c,k\ell}^{(i)}[m_1, m_2]$, $c=1,2,3$, are given in Section SM3. 

Software to estimate $\bm C_{k\ell}$ accurately and efficiently using the method described above is provided in the R package \texttt{concordance} which can be found at \url{https://github.com/knrumsey/concordance}. Using similar methods to the above, the software can also approximate the modification proposed by \cite{lee2019modified}, given in \cref{eq:modifiedasm} (see SM3 for details).  A numerical study of the Piston function \citep{zacks1998modern} is provided in Section SM4 in order to demonstrate the accuracy of the proposed approach for a moderately high-dimensional model.

\subsubsection{Simple polynomial example}
We will demonstrate the speed and accuracy of the described approach by using it for the simple polynomial example of \cref{eq:simple_example} with $\beta = 3$. We begin by sampling $n=200$ input locations using a maximin Latin hypercube, $\{\bm x_1, \bm x_2, \ldots, \bm x_n\}$ \citep{park1994optimalLHS}. For each of the generated inputs, we evaluate $f_1$ and $f_2$, and use the \texttt{BASS} R package to learn a representation of these functions in the form of \cref{eq:mars}. This takes $6.9$s and $7.2$s for $f_1$ and $f_2$ respectively, using a 2019 MacBook Pro with a 2.8 GHz Quad-Core Intel Core i7. Using these representations, we use \cref{eq:Capprox} to estimate $\bm C_1$, $\bm C_2$ and $\bm C_{12}$. Using the \texttt{C\_bass} and \texttt{Cfg\_bass} functions from the \texttt{concordance} package, an estimate was obtained for all $3$ matrices in just $0.233$ seconds. The estimated and true $\bm C_{12}$ matrices are 
\begin{equation}
\label{eq:simple_results2}
    \bm C^\text{true}_{12} \approx \begin{bmatrix}
    2.667 & 6.167 \\
    0.917 & 1.833
    \end{bmatrix}, \quad\quad\quad
    \bm C^\text{est}_{12} = \begin{bmatrix}
    2.634 & 6.086 \\
    0.896 & 1.790
    \end{bmatrix},
\end{equation}
with a Frobenius distance \citep{golub2013matrix} between the two matrices of $0.050$. If we increase the size of the training set to $n=1000$, the BASS models take between $21$ and $32$ seconds to fit, the matrices can be estimated with the \texttt{concordance} package in $1.14$ seconds, and the Frobenius distance decreases to $0.0086$. For a detailed scaling study and for details on the choice of prior $\mu(\bm x)$, see \cite{rumsey2023discovering}.

\section{Application to High Explosives}
\label{sec:HE}
We now return to the computer experiments for PBX 9501 described in \cref{sec:pbxintro}. The co-active subspace methods developed in \cref{sec:CASM} will be deployed in order to answer scientific questions of interest. The data and the R code used to generate many of the figures and quantities in this section can be found at \url{http://github.com/knrumsey/PBX9501}.

\begin{figure}
    \centering
    \begin{subfigure}[t]{0.48\textwidth}
        \centering
        \includegraphics[width=0.98\linewidth]{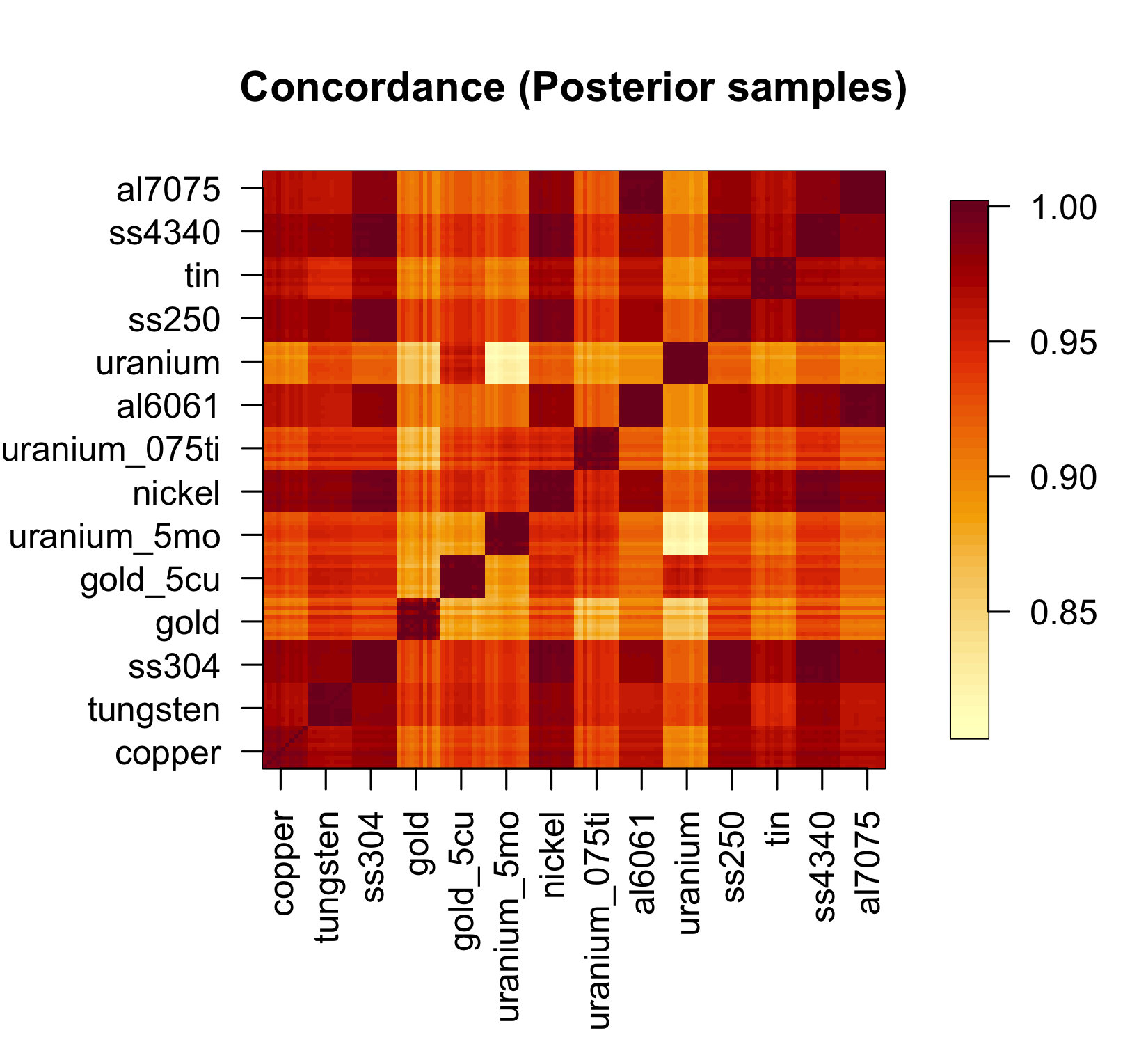}
        \caption{The estimated concordance for each of the $9370$ model pairs. The blocks represent posterior samples of the concordance estimate for pairs of jacket materials. }  
            \label{fig:conc_post}
    \end{subfigure}
    \begin{subfigure}[t]{0.48\textwidth}
        \centering
        \includegraphics[width=0.98\linewidth]{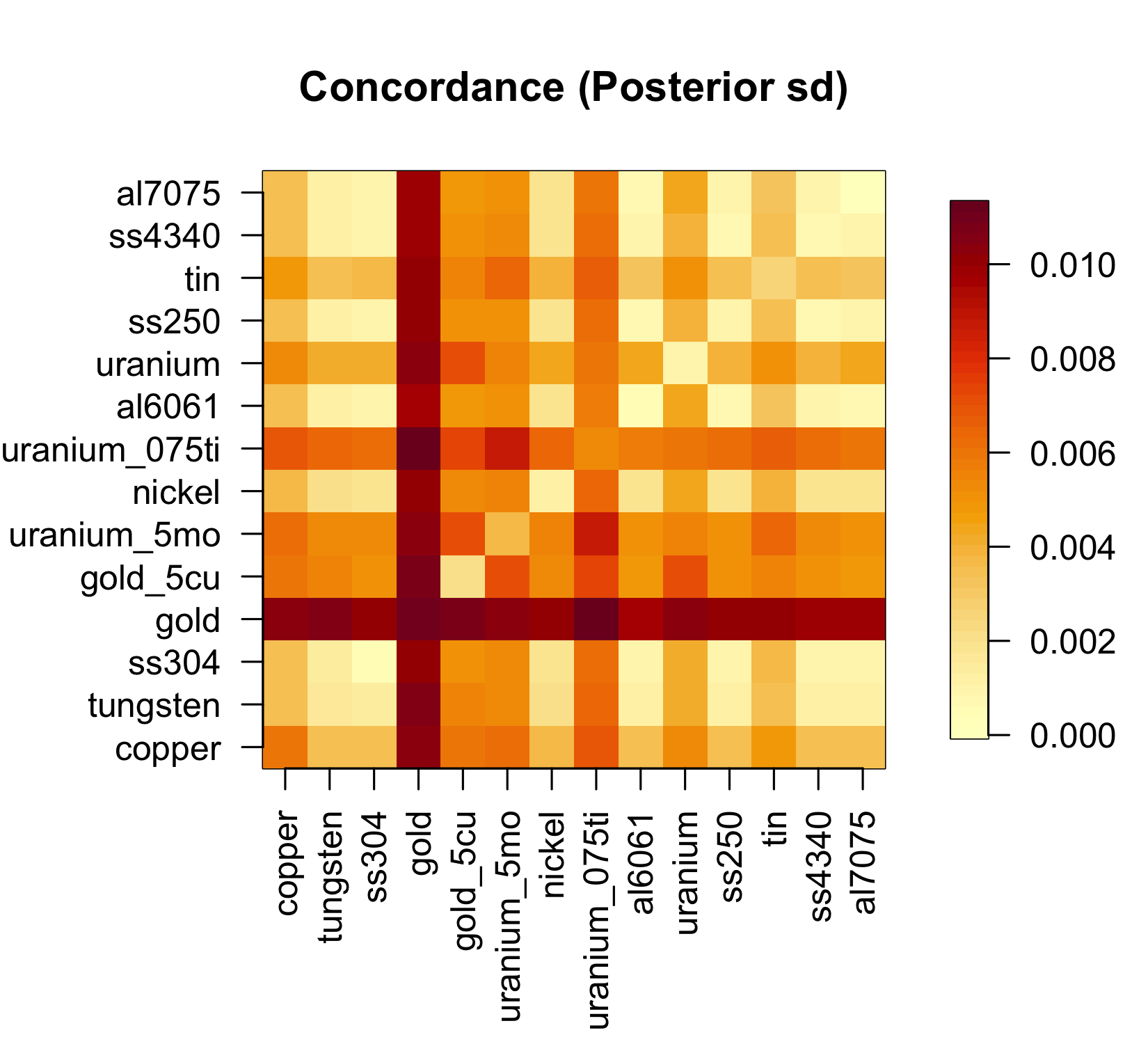}
        \caption{The posterior standard deviation for the pairwise concordance estimates. Even in the worst case, the posterior standard deviation is quite small. }
    \label{fig:conc_post_sd}
    \end{subfigure}
    \caption{}
\label{fig:concor}
\end{figure}

For each of the $K=14$ functions (corresponding to different jacket materials), the $500$ computer model runs are used to train a Bayesian MARS model, using the \texttt{BASS} package \citep{francom2020bass}, so that each $f_k$ can be represented in the form of \cref{eq:mars}. In each case, the BASS model required about $10$ seconds to learn the appropriate representation. The $14$ model fits were validated using $100$-fold cross validation (CV) \citep{stone1978cross}, where the CV root mean square prediction error (RMSPE) was found to be between $0.006$ and $0.108$ (the response was scaled to have unit variance). The worst model fits were obtained for gold, uranium and gold + 5\% copper.  

The posterior distribution of a BASS model can be viewed as an ensemble of MARS models. To account for uncertainty in the fits, we obtain $10,000$ ensemble members discarding the first $9,000$ for burn-in and thinning down to $10$ posterior samples (e.g., ensemble members) for each of the $14$ cases leading to a total of $140$ different functions of interest. The choice of ten ensemble members per computer model is arbitrary, and should be chosen according to the computational budget. In total, there are $\binom{14\times 10}{2} = 9730$ $\bm C_{k\ell}$ matrices that must be estimated (on top of the $140$ $\bm C_k$ matrices); a monumental task for MC based estimation. Fortunately, using the techniques developed in \cref{sec:implementation}, the \texttt{concordance} package is able to conduct the entire analysis in under half an hour on a 2019 MacBook Pro. For all $9730$ cases, the concordance is computed for the function pair using \cref{eq:conc}. 

Each computer model, corresponding to a particular jacket material, is represented in the analysis by $10$ functions. Thus for each pair of computer models (two distinct metals), there are $100$ function combinations and thus $100$ ``posterior draws" for the concordance. These samples can be seen in \cref{fig:conc_post}, where the underlying $14\times 14$ structure can easily be seen in the checkerboard pattern of each block. The overall estimate (posterior mean) of concordance for two metals (e.g., gold and copper), is obtained by averaging the values in the corresponding block. The posterior standard deviation, as seen in \cref{fig:conc_post_sd}, is calculated similarly. The relatively small posterior standard deviations indicate precise estimates, with the most uncertainty occurring when computing the concordances for gold. Explicit values for the posterior means and standard deviations are reported in Section SM5. 

\begin{figure}[b!]
    \centering
    \includegraphics[width=\linewidth]{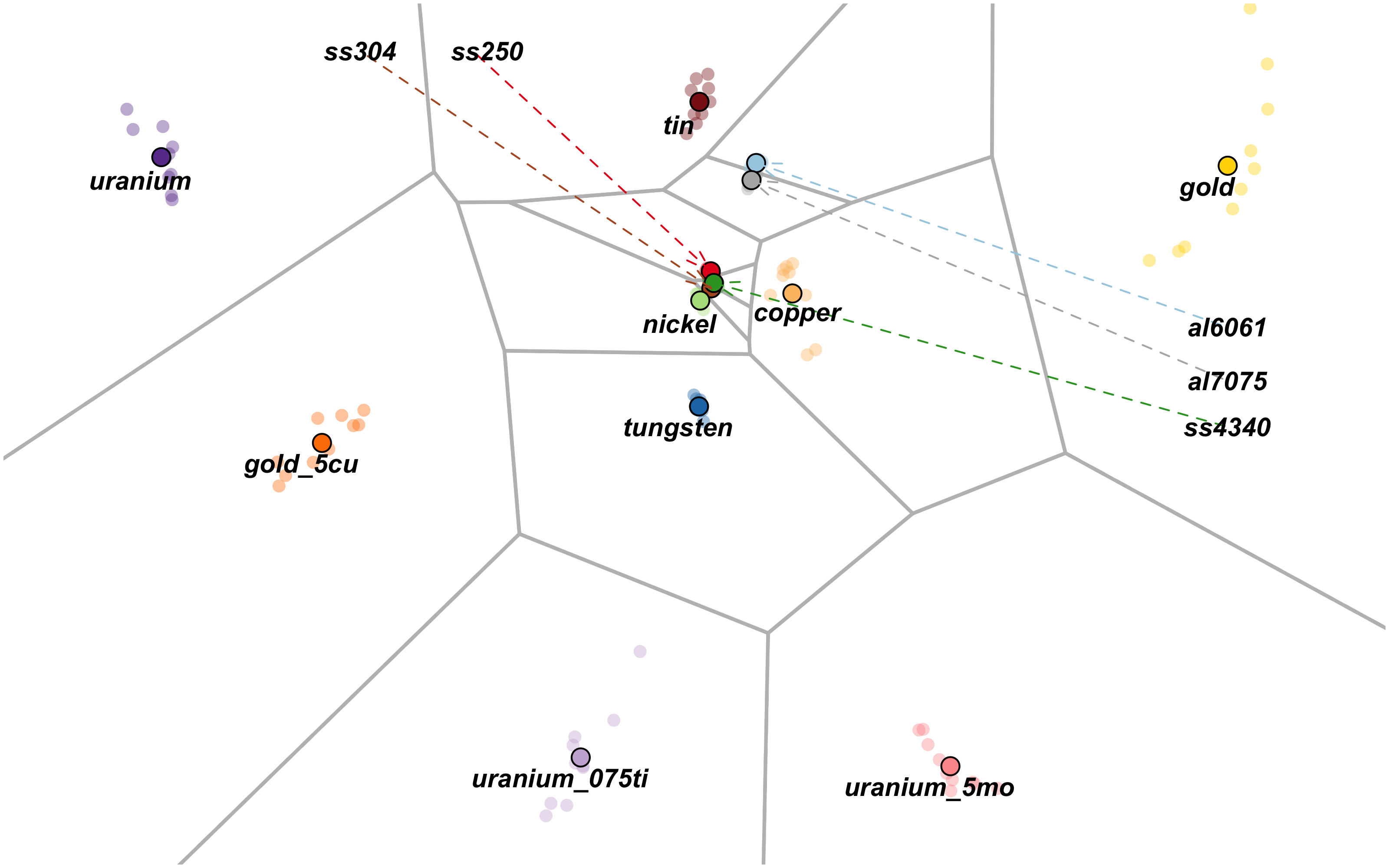}
    \caption{Clustering of jacket materials based on multidimensional scaling of discordance.}
    \label{fig:voronoi}
\end{figure}

At present, it is challenging (or at least time-consuming) to gain insight into things by simply staring at \cref{fig:concor}. Some interesting observations are apparent; all concordance values are positive and reasonably large, yet we notice the relatively low concordance ($0.820$) between the computer models with a uranium jacket and a uranium - 5\% Molybdenum jacket, which is surprising. General conclusions are hard to come by. To better illustrate these findings, we propose the following procedure: (i) convert the $140\times 140$ concordance matrix into a ``discordance matrix" by applying \cref{eq:discord} elementwise, (ii) use a multidimensional scaling (MDS) method (we choose Kruskal's non-metric MDS here; \cite{kruskal1964nonmetric, mead1992review}) to project the $140$ functions into a 2D space such that the Euclidean distance between points is approximately proportional to the discordance and (iii) define a point ``center" (e.g., first moment of the posterior point cloud) for each of the $14$ computer models and partition the 2D space using a Voronoi diagram \citep{okabe2009spatial}.

By following this approach, as seen in \cref{fig:voronoi}, the similarities and differences in the gradients of the various models become clearer and neighborhoods become easier to define and more intuitive to interpret. Each metal jacket is defined by a large number of system variables that define the EoS, strength, melt, and damage models. Although one can reasonably assume that some system variables, such as material density, strongly influence the velocity of the jacket at the PDV measurement location and time, the exact relationship between these variables and the resulting output is unknown, and likely quite complex. With this, it can be expected that a concordance analysis will bolster these underlying assumptions, and perhaps yield additional insights into the relationship between the system variables and PDV outputs. At the outset, it was assumed that density was the most determinate factor in the resultant PDV measurement. The denser the jacket, the slower it will move. Moderate density materials such as nickel, stainless steel, tin, and copper were expected to be more concordant with one another than with low densities, like aluminum, or high density materials, like uranium, tungsten, or gold. Upon initial investigation, \cref{fig:voronoi} seems to bear this out, however, there are some notable discrepancies within the high density materials. Namely, gold is more concordant with lower density materials than its high density counterparts. As it turns out, gold is a quite soft metal compared to others in the set, such as uranium. The hardness of a metal, measured by the work hardening parameter in \cref{tab:jacket_materials}, determines how resistant it is to deformation. A soft metal, therefore, may deform much more than a hard metal, resulting in notably different velocity measurements. Close inspection of \cref{tab:jacket_materials} and \cref{fig:voronoi} provide evidence for this claim. For example, tungsten, the softest of the high density jackets is more concordant with gold than any other. Further, as the hardness of the high density jackets increases, the concordance with gold decreases. It is clear that there is much more influencing the PDV output than merely the density and hardness, however, the concordance analysis not only confirmed the initial hypothesis of a strong correlation to density, but also highlighted an unexpected additional relationship between the system variables and the PDV output. We note that this example and the associated discussion are closely related to Q1 and Q2 in \cref{sec:adjacent}.

\subsection{Co-Sensitivities}
\label{sec:cosenspbx}
To continue the analysis, we will take a closer look at the co-active subspace and some related quantities. The questions we seek to answer in this subsection can be seen as an expanded version of Q3 from \cref{sec:adjacent}. For simplicity, we reduce the scope slightly and focus on just $3$ jacket materials: Stainless Steel 304 (SS), Nickel (Ni) and Uranium (U). From \cref{fig:voronoi} we can see that SS and Ni are neighbors in the sense that they are highly concordant ($\text{conc}(f_{SS}, f_{Ni}) = 0.996$), suggesting that they behave similarly under directional perturbation and are likely to be sensitive to the inputs in a similar way. Uranium, on the other hand, is much less concordant with the other two materials ($\text{conc}(f_{U}, f_{Ni}) = 0.921, \ \text{conc}(f_{U}, f_{Ni}) = 0.918$). In all cases, the active and co-active subspaces are effectively one-dimensional. 
\begin{table}[!htbp] \centering 
  \caption{The contribution vectors suggest that the majority of the co-activity is found in one direction.} 
  \label{tab:contrib} 
\begin{tabular}{@{\extracolsep{5pt}} lcccccc} 
\\[-1.8ex]\hline 
\hline \\[-1.8ex] 
& \multicolumn{6}{c}{Co-active direction} \\ \\[-1.8ex]
& $1$ & $2$ & $3$ & $4$ & $5$ & $6$ \\  \hline \\[-1.8ex]
$\bm\pi^{SS,Ni}$ & $0.995$ & $0.0004$ & $0.0004$ & $0.0002$ & $0.0001$ & $0.00002$ \\ 
$\bm\pi^{Ni,U}$ & $0.924$ & $0.001$ & $0.001$ & $0.0001$ & -$0.0003$ & -$0.004$ \\ 
$\bm\pi^{SS,U}$ & $0.922$ & $0.001$ & $0.0003$ & $0.00002$ & -$0.0002$ & -$0.005$ \\ 
\hline \\[-1.8ex] 
\end{tabular} 
\end{table} 
This can be seen by examination of the contribution vectors, reported in \cref{tab:contrib}, which show that the concordance can be almost entirely attributed to the first co-active direction. Since each $\pi_i$ is positive for the combination of SS and Ni, we learn that all active directions lead to concordant changes in the output of the computer models. The negative contributions $\pi_5$ and $\pi_6$ for the combination of U and SS (or equivalently U and Ni), indicate that moving along the $5^{th}$ and $6^{th}$ active direction has an opposite average-effect on the two functions. Note also that, for both combinations involving U, the $6^{th}$ contribution vector is the second largest in magnitude, and hence the corresponding active direction is actually the second most influential, though it is effectively negligible compared to the dominant direction. 

\begin{table}[!htbp] \centering 
  \caption{Activity and co-activity scores for Stainless Steel 304 (SS), Nickel (Ni) and Uranium (U). In all cases, the output of the computer model is most sensitive (or co-sensitive) to $r_1$ and least sensitive (or co-sensitive) to $\omega$.  } 
  \label{tab:activity} 
\begin{tabular}{@{\extracolsep{5pt}} lcccccc} 
\\[-1.8ex]\hline 
\hline \\[-1.8ex] 
& \multicolumn{6}{c}{Native input} \\ \\[-1.8ex]
 & $\rho_0$ & $A$ & $B$ & $r_1$ & $r_2$ & $\omega$ \\ 
\hline \\[-1.8ex] 
$10^3\times\bm\alpha^{SS}(1)$ & $5.968$ & $4.447$ & $3.835$ & $20.340$ & $6.922$ & $1.492$ \\[1.1ex] 
$10^3\times\bm\alpha^{Ni}(1)$ & $5.666$ & $4.578$ & $3.741$ & $20.506$ & $6.675$ & $1.362$ \\[1.1ex] 
$10^3\times\bm\alpha^{U}(1)$ & $4.158$ & $4.855$ & $2.799$ & $21.575$ & $4.122$ & $0.833$ \\[1.1ex]
$10^3\times\bm\alpha^{SS,Ni}(1)$ & $5.830$ & $4.532$ & $3.764$ & $20.397$ & $6.794$ & $1.448$ \\[1.1ex] 
$10^3\times\bm\alpha^{Ni,U}(1)$ & $4.561$ & $4.485$ & $3.312$ & $20.350$ & $5.495$ & $1.125$ \\[1.1ex] 
$10^3\times\alpha_{SS,U}(1)$ & $4.690$ & $4.420$ & $3.373$ & $20.266$ & $5.607$ & $1.150$ \\ 
\hline \\[-1.8ex] 
\end{tabular} 
\end{table} 

\begin{figure}
\label{fig:coactivity12}
    \centering
    \begin{subfigure}[t]{0.48\textwidth}
        \centering
        \includegraphics[width=0.98\linewidth]{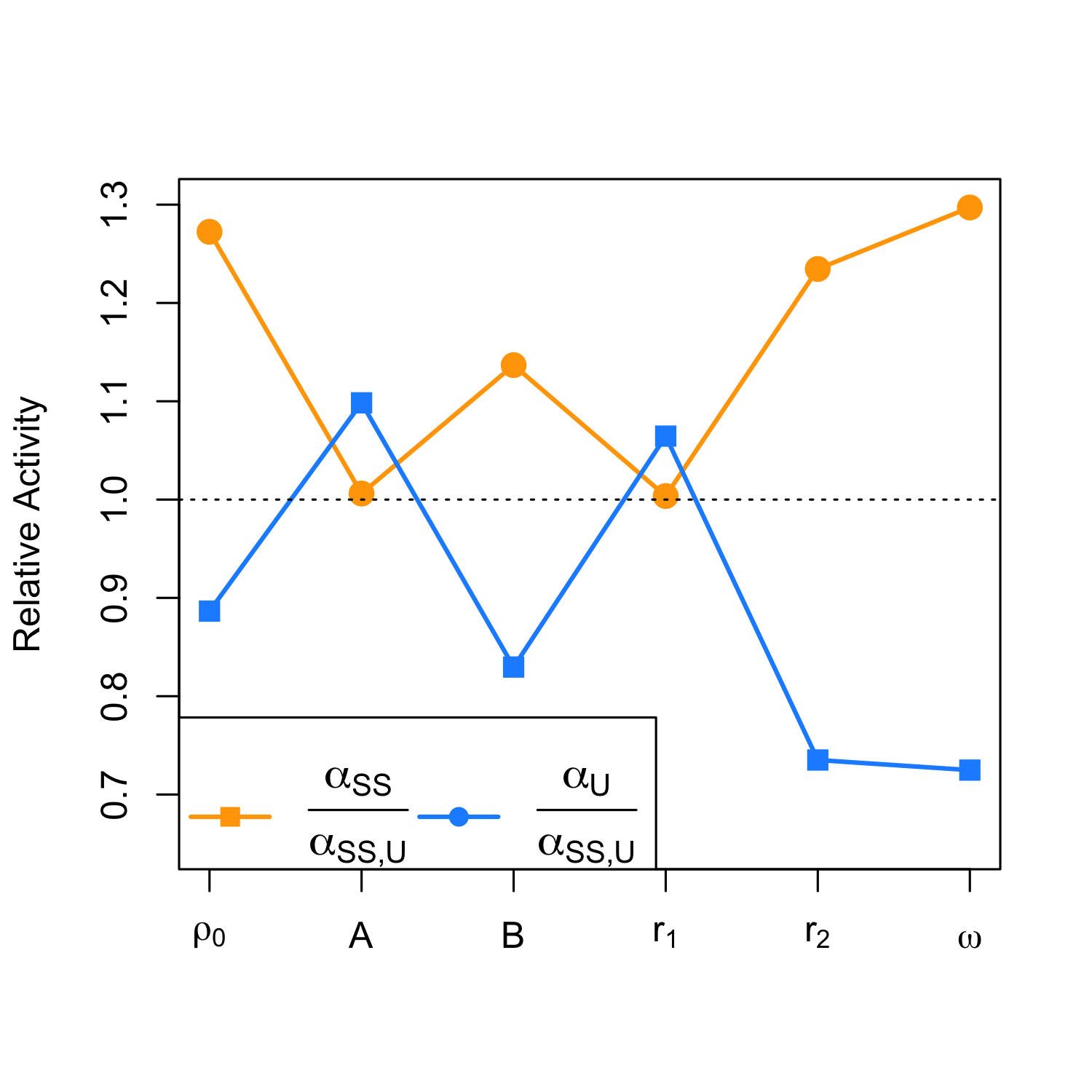}
        \caption{The relative activity of SS and U to their joint co-activity. We see that SS is much more sensitive to $\rho_0$, $B$, $r_2$ and $\omega$ than U. } 
    \label{fig:coactivity1}
    \end{subfigure}
    \begin{subfigure}[t]{0.48\textwidth}
        \centering
        \includegraphics[width=0.98\linewidth]{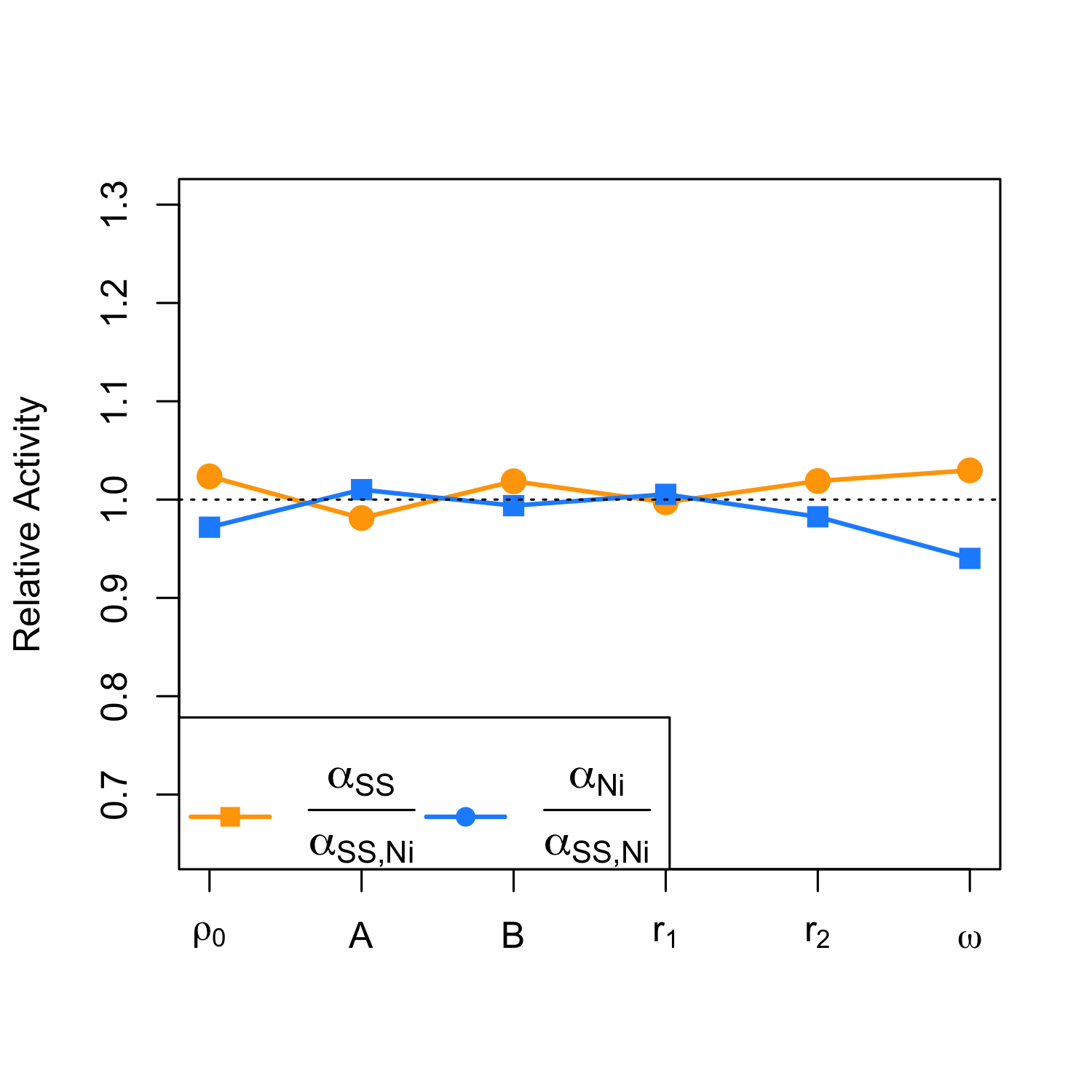}
        \caption{The relative activity of SS and Ni to their joint co-activity. We see that SS is more sensitive to $\rho_0$, $B$, $r_2$ and $\omega$ than Ni, but the difference is hardly of practical significance. }
    \label{fig:coactivity2}
    \end{subfigure}
    \caption{}
\end{figure}

Since the dominant (largest in magnitude) component of the contribution vector is positive, the signed and unsigned co-activity scores will be identical for the choice $q=1$. The activity and co-activity scores are reported in \cref{tab:activity}. Since all values are positive, we can deduce that the inputs affect the different computer models in the same ``direction". That is, increasing an input $x_i$ leads to an average increase (or decrease) in the output for both functions. The relative differences vary however, indicating that a particular JWL input is more ``important" for one function over the other. The ratios $\alpha_i^k/\alpha_i^{k\ell}$ and $\alpha_i^\ell/\alpha^{k\ell}$ $(i=1,\ldots, p)$ provide a useful visualization for understanding (i) which function is more sensitive to each input and (ii) the degree of difference between the functions. Two examples of this are given in \cref{fig:coactivity1} and \cref{fig:coactivity2}, where we see that SS is more sensitive to perturbations in $\rho_0$, $B$, $r_2$ and $\omega$ compared to both U and Ni, but the magnitude of the ratios are much more substantial when comparing to uranium. 

\subsection{Low Dimensional Representation of JWL}
\label{sec:lowD_JWL}
Finally, consider the problem of finding a suitable transformation of the JWL parameters for PBX 9501 with reduced dimension (i.e., Q4 in \cref{sec:adjacent}). One of the most valuable features of ASM is its ability to answer this question. The traditional active subspace is, however, tailored to a single function and may not be appropriate for two or more functions. The co-active subspace analysis yields a solution to this problem for the two-function case. To demonstrate, the $p=6$ dimensional input data is projected down to $2$ dimensions using (i) the first and seconds active directions of $f_{SS}$, (ii) the first and second active directions of $f_U$, and (iii) the first and sixth co-active directions of $f_U$ and $f_{SS}$ (recall that $\pi_6^{SS, U}$ has the second largest absolute contribution). Finally, we compare to the approach described in \cite{zahm2020gradient}, which is specifically designed for this purpose and reduces to projecting onto the first two eigenvectors of $\bm H = \bm C_{SS} + \bm C_U$.

\begin{table}[!htbp] \centering 
    \caption{Cross validation RMSPE values for BASS regression models for the PDV response (scaled to have unit variance), corresponding to Stainless Steel 304 (SS) and Uranium (U). The models were trained using various $2D$ projections of the input data. The projection onto the co-active subspace leads to accurate predictions for both materials simultaneously.}
    \label{tab:lower_dim}
\begin{tabular}{@{\extracolsep{5pt}} lccccc} 
\\[-1.8ex]\hline 
\hline \\[-1.8ex] 
 & Full Data & $\bm C_{SS}$ & $\bm C_U$ & $\bm V_{SS,U}$ & $\bm H$ \\ 
\hline \\[-1.8ex] 
Stainless Steel 304 & $0.011$ & $0.022$ & $0.145$ & $0.023$ & $0.038$ \\ 
Uranium & $0.074$ & $0.117$ & $0.078$ & $0.101$ & $0.097$ \\ 
\hline \\[-1.8ex] 
\end{tabular} 
\end{table} 

For each of these projected inputs, we fit non-linear surrogate models using the R \texttt{BASS} package \citep{francom2020bass} to predict the PDV velocity (see \cref{fig:setup}) in the experiments for Stainless Steel 304 and for Uranium. Shown in \cref{tab:lower_dim}, the quality of each fit is assessed by the $50$-fold cross validation RMSPE. The projected inputs designed specifically for one material led to good predictions for the intended material; the predictions are substantially less accurate when used to predict outcomes for the other material. The inputs projected onto the co-active subspace, on the other hand, lead to reasonably good predictions for both materials concurrently. Similar behavior is observed when projecting onto the $\bm H$ matrix of \cite{zahm2020gradient}, although the accuracy of the reduced model for Stainless Steel 304 is considerably worse when projecting onto $\bm H$ compared to $\bm V_{SS,U}$. We do acknowledge, however, that \cite{zahm2020gradient} could be used to construct a shared active subspace for all $14$ materials at once, while co-active subspaces (in their current form) are only useful for two functions at a time. 

\section{Conclusion}
\label{sec:conclusions}

In this paper, we have extended the active subspace methodology of \citep{constantine2015activebook} in a manner that not only aligns naturally with our problem but also introduces a novel dimension to this powerful tool. Other methods, such as the multivariate extension of \cite{zahm2020gradient} find shared active directions but are blind to whether the functions respond in a coordinated fashion along those directions.  Co-active subspace analysis captures the {\em sign} of correlated activity and degree of coordination shared by the two functions as they respond to input perturbations. 

As described in the introduction, co-activity is one of several approaches being used to demonstrate the functional equivalence of a new design with trusted, established systems. Our argument revolves around the high concordance observed between the new design and several well-established systems. This concordance suggests that the new design reacts similarly to input perturbations, and with further insights into the underlying physics, this congruence can help formulate a compelling case that the same dominant physical phenomena are at play. It is important to recognize that this work is not exhaustive, and further refinement of this argument is warranted. One avenue we propose is the use of the discordance matrix across a collection of related systems. This approach enables us to quantitatively assess the proximity of each system to the new design and explore the possibility of partitioning the systems into meaningful clusters. The quantitative measure of similarity between the new design and trusted systems plays a pivotal role in bolstering our overall confidence in the new design, particularly when the feasibility of full-system testing is constrained.

Looking ahead, there are promising avenues for extension. The current framework, which focuses on pairwise comparisons, could be further developed to accommodate more than two functions, enabling a more comprehensive analysis. Additionally, temporal or spatial variations could be explored to address the nuances of dynamic systems. 

Despite its utility, we have noted the limitations that arise from symmetrizing the $\bm C_{fg}$ matrix. Future work could focus on studying $\bm C_{fg}$ directly, perhaps through its singular value decomposition, in the pursuit of a fully comprehensive analysis. Finally, {\em active manifolds} \citep{bridges2019active} extend AS methods to find nonlinear streamline curves through the input space along which the output maximally varies.  Conceptually, a co-active manifold would find a curve along which the gradients of two adjacent models are maximally aligned. These extensions hold the potential to expand the applicability of our methodology to a wider array of complex engineering challenges.

\section*{Supplemental Materials}
The supplemental materials include additional details for the rate stick experiments, a proof that discordance satisfies the properties of a pseudo-metric, implementation details (including a brand new derivation of the expected gradient in closed-form for models in the form of \cref{eq:mars}), a higher-dimensional example using the Piston function \citep{zacks1998modern}, and posterior means and standard deviations corresponding to \cref{fig:coactivity12}. An R package is also available and can be found at \url{https://github.com/knrumsey/concordance}.

\bibliographystyle{agsm}
\bibliography{references}

\newpage
\pagebreak

\begin{center}
\Large Supplemental Materials
\end{center}

\titleformat{\section}{\normalfont\Large\bfseries}{SM\thesection}{1em}{}
\setcounter{section}{0}

\section{Additional Details on Rate Stick Experiment Diagnostics}

This brief section goes into additional details regarding the simulated PDV probes. 
PDV probes function by measuring the reflected light from a laser directed at a surface.
When the surface is moving, this will result in a small shift in the the laser frequency, which can be directly related to the velocity of the surface.

In each of the simulations, five different PDV probes were modeled with equidistant axial spacing.
Computationally, the probes are defined with by their origin and laser direction.
The probes are defined 3.9 cm from the outer metal jacket surface (5 cm from the origin) with 1.6 cm axial spacing, and are directed along the radial axial (directly at the outer surface).

\begin{figure}[htp]
    \centering
    \includegraphics[width=0.8\linewidth]{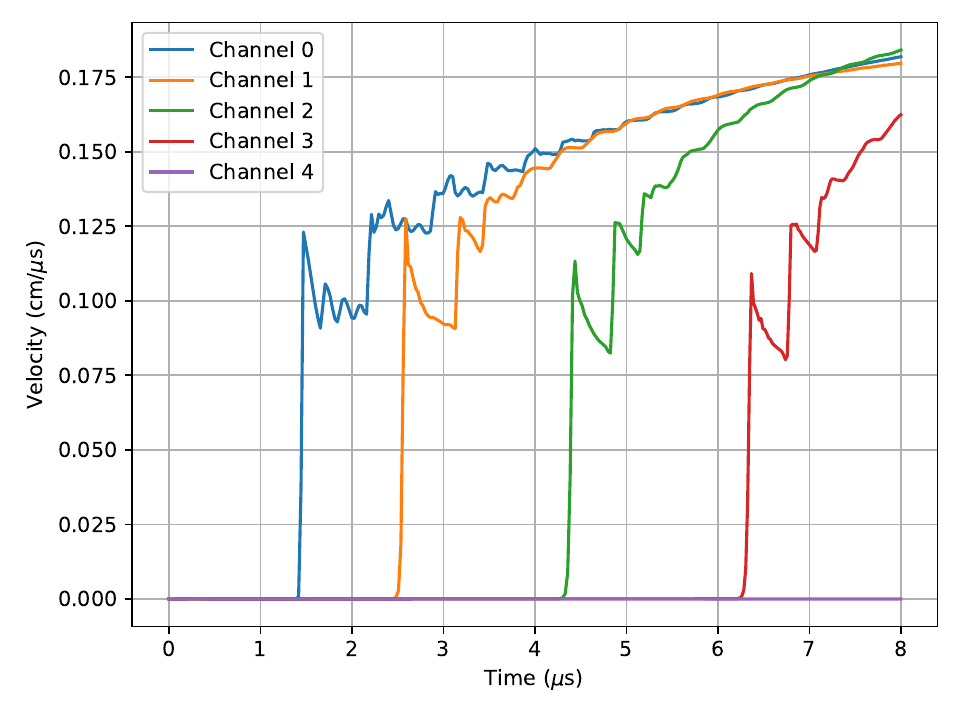}
    \caption{The PDV signal from a nominal rate stick experiment with a copper jacket.} 
    \label{fig:pdv_profiles}
\end{figure}

\cref{fig:pdv_profiles} shows the time-dependent PDV data for each of the probes.
The initial rise in the velocity profiles indicates the arrival time of the shock at the outer boundary of the problem at the respective probe elevations.
From the spacing in the shock arrival times, it is clear that the speed of the detonation front changes throughout the simulation.
Once the detonation front passes channel 1, the detonation speed remains relatively constant, indicating the detonation front has reached its steady burn rate.
This characteristic provided the primary rationale in choosing to analyze PDV measurements at later times and higher axial positions.

Several more important features can be taken from \cref{fig:pdv_profiles}.
The velocity profiles exhibit quite complex behaviors.
Once the initial shock has reached the outer surface of the metal jacket, a rarefaction wave is reflected back into the jacket.
This leads to the subsequent decrease in velocity.
The remaining behaviors are explained by the shocks ``ringing`` in the system.
Each time a shock reaches an interface, part of it is transmitted to the neighboring material, and part is reflected as either a shock or rarefaction wave.
As time goes on, the strength of the shock is attenuated, but many more wave fronts exist for a give elevation.
This leads to the velocity profiles smoothing out over time.
Because the aim was to compare the concordance of different metal jackets, it was important to choose a measurement far enough away from the shock front arrival time to capture the integral material response for comparison.

\section{Discordance as a pseudo-metric} 
A pseudo-metric is a metric but without requiring the identity of indiscernibles,  meaning two objects with zero distance are necessarily the same. This section demonstrates that discordance defined in Eq. (10) is a pseudo-metric.
Functions $f_1, f_2, \ldots$ are taken to have inputs $\bm x$.  
Discordance is defined as 
\begin{equation}
\label{eq:discord2}
    \text{discord}(f_k, f_{\ell}) = \sqrt{\frac{1-\text{conc}(f_k, f_{\ell})}{2}},
\end{equation}
with concordance defined in Eq. (9).

The four properties of a pseudo metric are:
\begin{description}
    \item[non-negative:] $d(f_k,f_{\ell}) \geq 0$ follows from $\text{conc}(f_k,f_{\ell}) \in [-1,1]$.
    \item[zero self-distance:] $d(f_k,f_k) = 0$ follows from Eq. (9) and (10) by recalling that $t_{kk} = t_k$ so $\text{conc}(f_k,f_k)=1$.
    \item[symmetry:] $d(f_k,f_{\ell}) = d(f_k,f_{\ell})$ follows from Eq. (9) by noting from Eq. (5) that $t_{k\ell}=t_{\ell k}$
\end{description}
The final property is the {\bf triangle inequality}
\[ d(f_k,f_{\ell}) \leq d(f_k,f_m) + d(f_m,f_{\ell}) \]
which requires a little work. For an i.i.d.~sample 
$\bm x_1, \bm x_2, \ldots \sim \mu$ 
the Strong Law of Large Numbers gives
\begin{equation} \label{eq:sampleTrace}
    t_{k\ell}^{(L)} \equiv
    L^{-1} \sum_{m=1}^L \nabla f_k(\bm x_i)^\intercal \nabla f_\ell(\bm x_i) 
    \overset{\text{a.s.}}{\longrightarrow}
    E_\mu\left[ \nabla f_k(\bm x)^\intercal \nabla f_\ell(\bm x) \right]
    = t_{k\ell}
\end{equation}
as $L \rightarrow \infty$.
Define the vector
\[ \bm z_{kL}
    = \left(\nabla f_k(\bm x_1)^\intercal, \ldots, \nabla f_k(\bm x_L)^\intercal \right)^\intercal
    \]
so that (assuming gradients are not a.s. zero)
\begin{equation} \label{eq:ckl}
    \kappa_{k\ell}^{(L)}
    \equiv
    \frac{\bm z_{kL}^\intercal \bm z_{\ell L}}
        {\sqrt{
        \left( \bm z_{kL}^\intercal \bm z_{kL} \right)
        \left( \bm z_{\ell L}^\intercal \bm z_{\ell L} \right) }}
        =
    \frac{t_{k\ell}^{(L)}}{\sqrt{t_{kk}^{(L)} t_{\ell\ell}^{(L)}}}
     \overset{\text{a.s.}}{\longrightarrow}
    \frac{t_{k\ell}}{\sqrt{t_{kk} t_{\ell\ell}}}
    = \text{conc}(f_k, f_\ell). 
\end{equation}
Thus, if $\sqrt{1-c_{\cdot\cdot}^{(L)}}$ obeys the triangle inequality, then so does $\text{discord}(\cdot,\cdot)$.
The definition of $c_{k\ell}^{(L)}$ in \cref{eq:ckl} shows it as the cosine of the angle between $\bm z_{kL}$ and $\bm z_{\ell L}$ and  
\citep[Sec.~4]{van2012metric} proves that $\sqrt{1-c}$ obeys the triangle inequality when $c$ is the cosine of the angle between Euclidean vectors.  

\noindent $\square$

\section{Details for efficient approximation of the co-Constantine matrix}

With slightly more generality than in the main text, suppose that $f^{(k)}(\bm x)$ can be written as
\begin{equation}
\label{eq:hfun}
    f_k(\bm x) \approx \gamma_{k0} + \sum_{m=1}^{M_k} \gamma_{km}\prod_{i=1}^ph^{(k)}_{im}(x_i),
\end{equation}
In this setting, the univariate integrals (from Eq. (23)) that we need to compute can be written as
\begin{equation}
    \begin{aligned}
        I_{1,k\ell}^{(i)}[m_1, m_2] &= \int_{-\infty}^\infty \frac{dh^{(k)}_{im_1}(x)}{d x}h^{(\ell)}_{im_2}(x) \mu_i(x) dx \\
        I_{2,k\ell}^{(i)}[m_1, m_2] &= \int_{-\infty}^\infty h^{(k)}_{im_1}(x) h^{(\ell)}_{im_2}(x) \mu_i(x) dx \\
        I_{3,k\ell}^{(i)}[m_1, m_2] &= \int_{-\infty}^\infty \frac{dh^{(k)}_{im_1}(x)}{d x}\frac{dh^{(\ell)}_{im_2}(x)}{d x} \mu_i(x) dx.
    \end{aligned}
\end{equation}
It is important to note that $I^{(i)}_{r,k \ell}$ and $I^{(i)}_{r,\ell k}$ are the same for $r = 2$ and $r = 3$, but these quantities must be computed separately for $r=1$. Additionally, none of the matrices are likely to be symmetric (or even square). This why (for $M=M_1=M_2$) there are $4pM^2$ univariate integrals to compute here, compared to just $pM(2M-1)$ integrals in the single function case explored by \cite{rumsey2023discovering}. 

In the case where \cref{eq:hfun} is a MARS model (as is the case for the \texttt{concordance} R package), we have 
$$h_{im}^{(k)}(x_i) = \left[\chi_{kim}(x_i)s_{kim}(x_i - t_{kim})\right]^{u_{kim}}$$
and
$$\frac{d }{dx_i}h_{im}^{(k)}(x_i) = u_{kim}s_{kim}\chi_{kim}(x_i)$$
where $\chi_{kim}(x) = \mathbbm{1}(s_{kim}(x_i-t_{kim} > 0))$. Thus, to compute the integrals above for this special case, we only need to evaluate truncated moments (with respect to $\mu_i$) of order $r \in \{0,1,2\}$
$$\xi(r | a, b, \mu_i) = \int_a^b x^r \mu_i(x) dx.$$
From here, letting $\xi(r)$ denote $\xi(r | a_{k\ell}^{(i)}[m_1, m_2], b_{k\ell}^{(i)}[m_1, m_2], \mu_i)$, it is straightforward to show that
\begin{equation}
\label{eq:I1mars}
I_{1,k\ell}^{(i)}[m_1, m_2] = s_{kim_1}s_{\ell im_2}
\begin{dcases}
u_{kim_1}\left(\xi(1) - t_{\ell im_2}\xi(0)\right), & u_{\ell im_2} = 1 \\[1.5ex]
u_{kim_1}\xi(0) , & u_{\ell im_2} = 0
\end{dcases} 
\end{equation}
\begin{equation}
\label{eq:I2mars}
I_{2,k\ell}^{(i)}[m_1, m_2] = s_{kim_1}s_{\ell im_2}
\begin{dcases}
\xi(2) - (t_{kim_1} + t_{\ell im_2})\xi(1) + t_{kim_1}t_{\ell im_j}\xi(0), & u_{kim_1} = 1, \ u_{\ell im_2} = 1 \\[1.5ex]
\xi(1) - t_{kim_1}\xi(0), & u_{kim_1} = 1, \ u_{\ell im_2} = 0 \\[1.5ex]
\xi(1) - t_{\ell im_2}\xi(0), & u_{kim_1} = 0, \ u_{\ell im_2} = 1 \\[1.5ex]
1, & u_{kim_1} = 0, \ u_{\ell im_2} = 0 
\end{dcases} 
\end{equation}
\begin{equation}
\label{eq:I3mars}
I_{3,k\ell }^{(i)}[m_1, m_2] = u_{kim_1}u_{\ell im_2}s_{kim_1}s_{\ell im_2}\xi(0). 
\end{equation}
The bounds, $a_{k\ell}^{(i)}[m_1, m_2], b_{k\ell}^{(i)}[m_1, m_2], \mu_i)$, are given by
\begin{equation}
\begin{aligned}
    \label{eq:bounds}
    a_{k\ell }^{(i)}[m_1, m_2] &= \begin{dcases}
    \max(t_{kim_1}, t_{\ell im_2}), & s_{kim_1} = +1, \ s_{\ell im_2} = +1 \\
    t_{kim_1}, & s_{kim_1} = +1, \ s_{\ell im_2} = -1 \\
    t_{\ell im_2},  & s_{kim_1} = -1, \ s_{\ell im_2} = +1 \\
    -\infty, & s_{kim_1} = -1, \ s_{\ell im_2} = -1
    \end{dcases} 
    \\[1.5ex]
    b_{\star,k\ell}^{(i)}[m_1, m_2] &= \begin{dcases}
    \infty, & s_{kim_1} = +1, \ s_{\ell im_2} = +1 \\
    t_{\ell im_2}, & s_{kim_1} = +1, \ s_{\ell im_2} = -1 \\
    t_{kim_1}, & s_{kim_1} = -1, \ s_{\ell im_2} = +1 \\
    \min(t_{kim_1}, t_{\ell im_2}), & s_{kim_1} = -1, \ s_{\ell im_2} = -1
    \end{dcases} \\[1.5ex]
    b_{k\ell}^{(i)}[m_1, m_2] &= \max\left\{b_{\star,k\ell}^{(i)}[m_1, m_2], a_{k\ell}^{(i)}[m_1,m_2]\right\}.
\end{aligned}
\end{equation}
This has been implemented in the \texttt{concordance} package. R code to conduct a simple toy analysis is given below. 

\begin{verbatim}
    #install.packages("BASS")
    #devtools::install_github("knrumsey/concordance")
    f <- function(x, scale=0) x[1]^2 + x[1]*x[2] + x[2]^3*scale
    X <- matrix(runif(300), ncol=3)
    y1 <- apply(X, 1, f)
    y2 <- apply(X, 1, f, scale=1/2)
    mod1 <- BASS::bass(X, y1)
    mod2 <- BASS::bass(X, y2)
    C <- concordance::Cfg_bass(mod1, mod2)
\end{verbatim}

\subsection{The Modified Co-ASM}
To compute the modified Co-ASM as suggested by \cite{lee2019modified} for quadratic functions (see Eq. (9) in the main text), we need to be able to compute the expected gradient
\begin{equation}
    Z_k = E[\nabla f_k(\bm x)].
\end{equation}
Using the same infrastructure as the above, we can write that
\begin{equation}
    E\left[\frac{\partial f_k(\bm x)}{\partial x_i}\right] = \sum_{m=1}^M\gamma_m I_{4,k}^{(i)}[m]\prod_{j\neq i}I_{5,k}^{(i)}[m].
\end{equation}
where $I_{4,k}^{(i)}[m] = \int_{-\infty}^\infty {h^{(k)}}^\prime_{im}(x) \mu_i(x) dx$ and $I_{5,k}^{(i)}[m] = \int_{-\infty}^\infty h^{(k)}_{im}(x) \mu_i(x) dx$. In the case of MARS, these integrals are given by 
\begin{equation}
    \begin{aligned}
        I_{4,k}^{(i)}[m] &= s_{kim}u_{kim}\xi(0|a_{kim}, b_{kim}, \mu_i) \\
        I_{5,k}^{(i)}[m] &= \left(s_{kim}\left(\xi(1|a_{kim}, b_{kim}, \mu_i) - t_{kim}\right)\xi(0|a_{kim}, b_{kim}, \mu_i)\right)^{u_{kim}} \\
        a_{kim} &= \frac{2s_{kim}t_{kim}}{s_{kim} + 1} \\
        b_{kim} &= \frac{2s_{kim}t_{kim}}{s_{kim} - 1}, 
    \end{aligned}
\end{equation}
where $a_{kim}, b_{kim} \in \mathbb R \cup \{-\infty, \infty\}$. The expected gradient vector can be computed with the \texttt{concordance} package as
\begin{verbatim}
    #install.packages("BASS")
    #devtools::install_github("knrumsey/concordance")
    f <- function(x, scale=0) x[1]^2 + x[1]*x[2] + x[2]^3*scale
    X <- matrix(runif(300), ncol=3)
    y1 <- apply(X, 1, f)
    mod1 <- BASS::bass(X, y1)
    Z <- concordance::Z_bass(mod1)
\end{verbatim}
An example can be found at \url{inst/CoASM/Z_bass_example.R} in the github repo. 

\subsection{Pseudocode}
Pseudo code for estimating the co-Constantine matrix $C_{12}$ for functions $f_1$ and $f_2$. The inputs to the algorithm include a set of inputs $\bm x_1, \ldots, \bm x_n$ and two sets of outputs $y_{1i} = f_1(\bm x_i)$ and $y_{2i} = f_2(\bm x_i)$ for $i=1,\ldots, n$. We also have to specify the (marginally independent) prior for each input $\mu_1, \ldots, \mu_p$. 
\begin{algorithm}[H]
    \SetAlgoLined
    \KwIn{$\mathcal (\bm x_1, y_{11}, y_{21}), \ldots, (\bm x_n, y_{1n}, y_{2n})$, $\mu_1, \ldots \mu_p$}
    \KwOut{Co-Constantine matrix: $\bm C$}
    $\mathcal M_1 \gets \text{BASS}(\{\bm x_1,\ldots, \bm x_n\}, \{y_{11}, \ldots, y_{1n}\})$ \tcp*{Fit emulators with \cite{francom2020bass}}
    $\mathcal M_2 \gets \text{BASS}(\{\bm x_1,\ldots, \bm x_n\}, \{y_{21}, \ldots, y_{2n}\})$ 
    
    {\bf Initialize:} $a, b, I_1, I_2, I_3 \in \mathbb R^{(\mathcal M_1.M) \times (\mathcal M_2.M) \times p}$
    
    \For{$m_1 \gets 1$ \KwTo $\mathcal M_1.M$}{
        \For{$m_2 \gets 1$ \KwTo $\mathcal M_2.M$}{
            \For{$i \gets 1$ \KwTo $p$}{
                \text{params} $\gets (\mathcal M_1.s, \mathcal M_1.t, \mathcal M_1.u, \mathcal M_2.s, \mathcal M_2.t, \mathcal M_2.u)$
                
                {\bf Compute } $a^{(i)}[m_1, m_2](\text{params})$ \tcp*{\Cref{eq:bounds}}
                                
                {\bf Compute } $b^{(i)}[m_1, m_2](\text{params})$ \tcp*{\Cref{eq:bounds}}

                {\bf Compute } $I_1^{(i)}[m_1, m_2](\text{params}, a^{(i)}[m_1, m_2], b^{(i)}[m_1, m_2], \mu_i)$ \tcp*{\Cref{eq:I1mars}}

                {\bf Compute } $I_2^{(i)}[m_1, m_2](\text{params}, a^{(i)}[m_1, m_2], b^{(i)}[m_1, m_2], \mu_i)$ \tcp*{\Cref{eq:I2mars}}

                {\bf Compute } $I_3^{(i)}[m_1, m_2](\text{params}, a^{(i)}[m_1, m_2], b^{(i)}[m_1, m_2], \mu_i)$ \tcp*{\Cref{eq:I3mars}}
            }
        }
    }

    {\bf Initialize} $\bm C_{12}$
    
    \For{$i \gets 1$ \KwTo $p$}{
        \For{$j \gets 1$ \KwTo $p$}{
        
            {\bf Compute } $\bm C_{12}[i,j](I_1, I_2, I_3, \mathcal M_1.\text{coeff}, \mathcal M_2.\text{coeff})$ \tcp*{Equation (25) in main text}
        }
    }

    {\bf Return } $\bm C_{ij}$
    \caption{Pseudocode for the computation of the co-constantine matrix $\bm C_{ij}$. We use the notation $\mathcal M.\text{field}$ to denote a ``field" or ``attribute" belonging to object (model) $\mathcal M$. }
    \label{algo:example}
\end{algorithm}

\section{Numerical study with Piston function}
In this example, we demonstrate the correctness of the numerical method proposed in the main manuscript by comparing the result to a Monte Carlo based approach. In most practical problems, the MC based approach suffers from limitations, as described in the main problem. Here, we use a simple analytic example for which the MC approach can be computed reasonably quickly, and thus the MC approximation for $\bm C_{12}$ is used as ground truth. 

\begin{table}[b]
    \centering
\begin{tabular}{l| c|c|c}
    notation & input & range & description \\ \hline
    $x_1$ & $M$ & $[30, 60]$ & piston weight (kg) \\
    $x_2$ & $S$ & $[0.005, 0.020]$ & piston surface area ($m^2$) \\
    $x_3$ & $V_0$ & $[0.002, 0.010]$ & initial gas volume ($m^3$) \\
    $x_4$ & $k$ & $[1000, 5000]$ & spring coefficient (N/m) \\
    $x_5$ & $T_0$ & $[340, 360]$ & filling gas temperature (K) \\
    $\xi_1$ & $P_0$ & $[90000, 110000]$ & atmospheric pressure (N/$m^2$) \\
    $\xi_2$ & $T_a$ & $[290, 296]$ & ambient temperature (K) \\
\end{tabular}
    \caption{Description of input variables and their ranges for the piston simulation function.}
    \label{tab:piston}
\end{table}

Consider the {\it Piston function} \citep{zacks1998modern, ben2007modeling}, traditionally defined as
\begin{equation}
    \label{eq:piston}
    \begin{aligned}
    f(\bm x) &= 120\pi \sqrt{\frac{M}{k+S^2\frac{P_0V_0T_a}{T_0V^2}}}, \quad\text{where} \\
    V &= \frac{S}{2k}\left(\sqrt{A^2+4k\frac{P_0V_0T_a}{T_0}}-A\right) \quad\text{and}\quad 
    A = P_0S + 19.62M - \frac{kV_0}{S},
    \end{aligned}
\end{equation}
with standard parameter ranges given in \cref{tab:piston}. We consider modeling the piston function for two different settings of the ambient conditions. The table indicates the $p=5$ global simulation variables ($x_1, \ldots, x_5$) and the simulation variables $\xi_1, \xi_2$. We define two computer models as
\begin{equation}
\begin{aligned}
    &f_1(\bm x | \xi_1=90000, \xi_2 = 284) \\ 
    &f_2(\bm x | \xi_1=110000, \xi_2 = 302),
\end{aligned}
\end{equation}
where the inputs $\bm x$ are scaled to $[0, 1]^5$ for simplicity. The matrix $\bm C_{12}$ is first estimated using a highly-accurate MC approach with $M=100,000$ draws and using forward automatic differentiation \citep{rall1996introduction} for gradient estimation (using the \texttt{fd\_grad()} function in the \texttt{concordance} package). For comparison, we also estimate the $\bm C_{12}$ matrix using the method described in Section 3.4 of the main manuscript (using the \texttt{C\_bass} function in the \texttt{concordance} package). R code to reproduce this example can be found at \url{http://github.com/knrumsey/concordance/inst/CoASM/piston\_coactive.R}. 

\begin{figure}[h!]
    \centering
        \begin{subfigure}[t]{0.31\textwidth}
        \centering
        \includegraphics[width=0.95\linewidth]{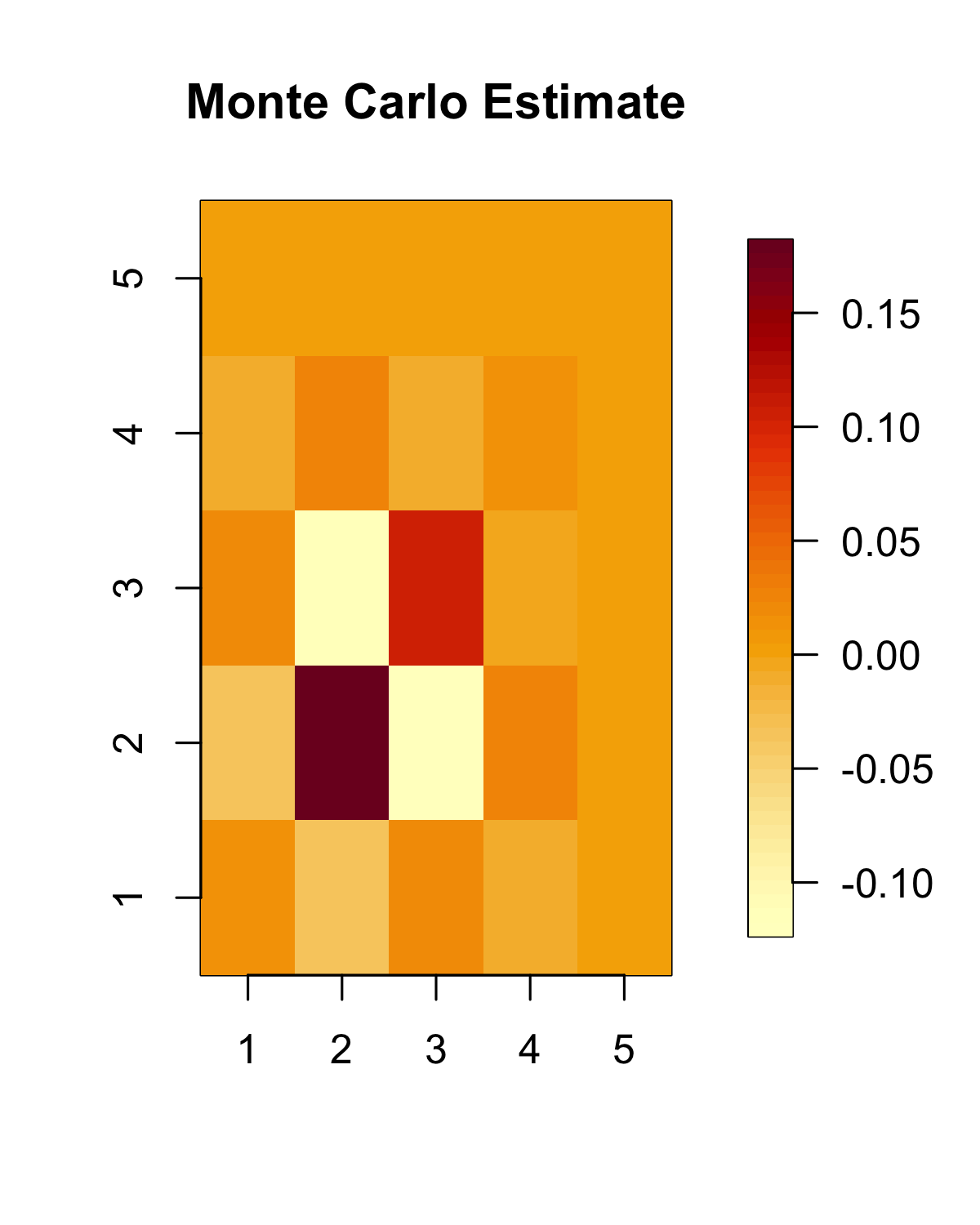}
        \caption{Monte Carlo estimate of $\bm C_{12}$ for the piston functions.}
    \label{fig:piston_mc}
    \end{subfigure}
    \begin{subfigure}[t]{0.31\textwidth}
        \centering
        \includegraphics[width=0.95\linewidth]{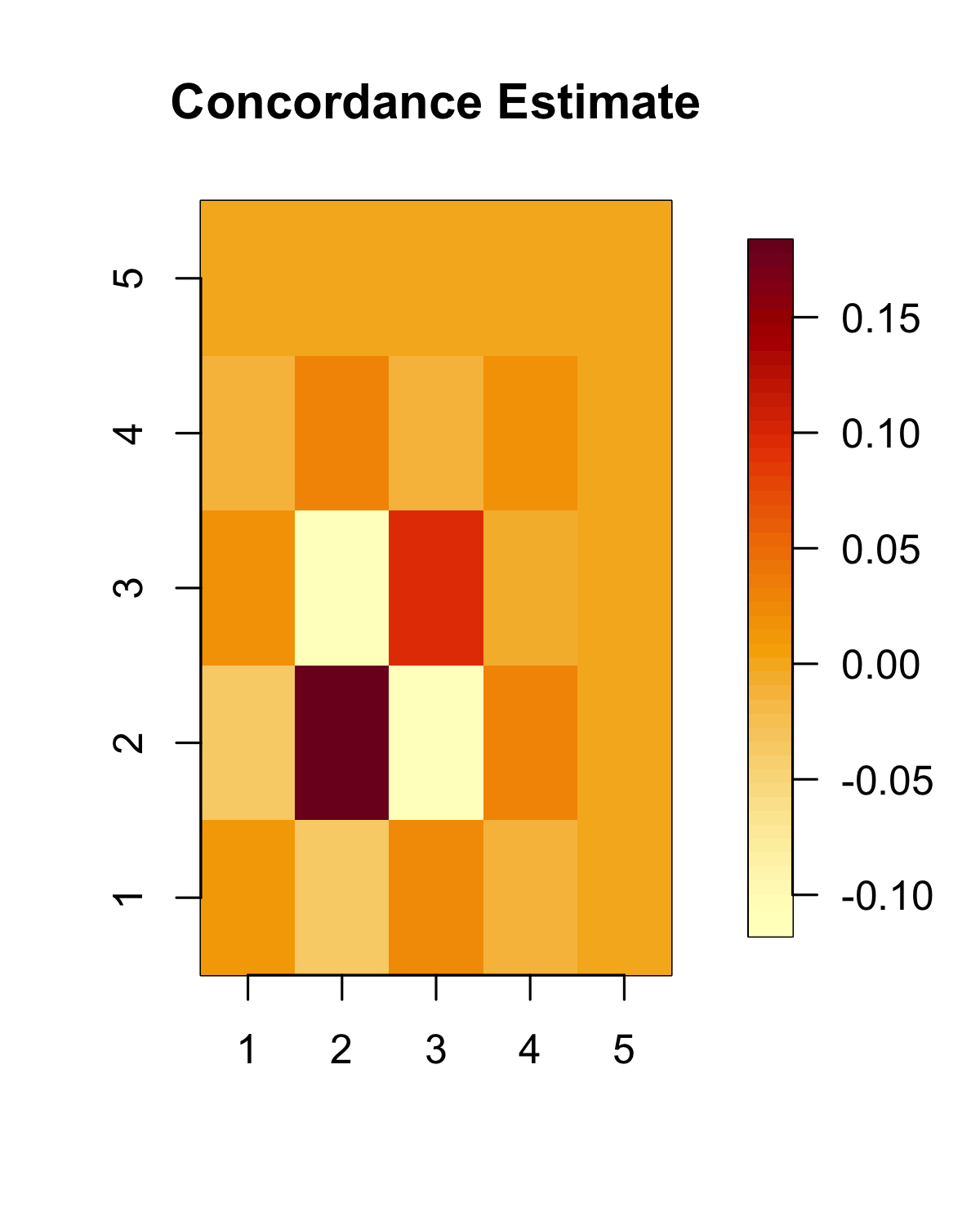}
        \caption{Concordance package estimate of $\bm C_{12}$ for the piston functions.}
    \label{fig:piston_conc}
    \end{subfigure}
    \begin{subfigure}[t]{0.31\textwidth}
        \centering
        \includegraphics[width=0.95\linewidth]{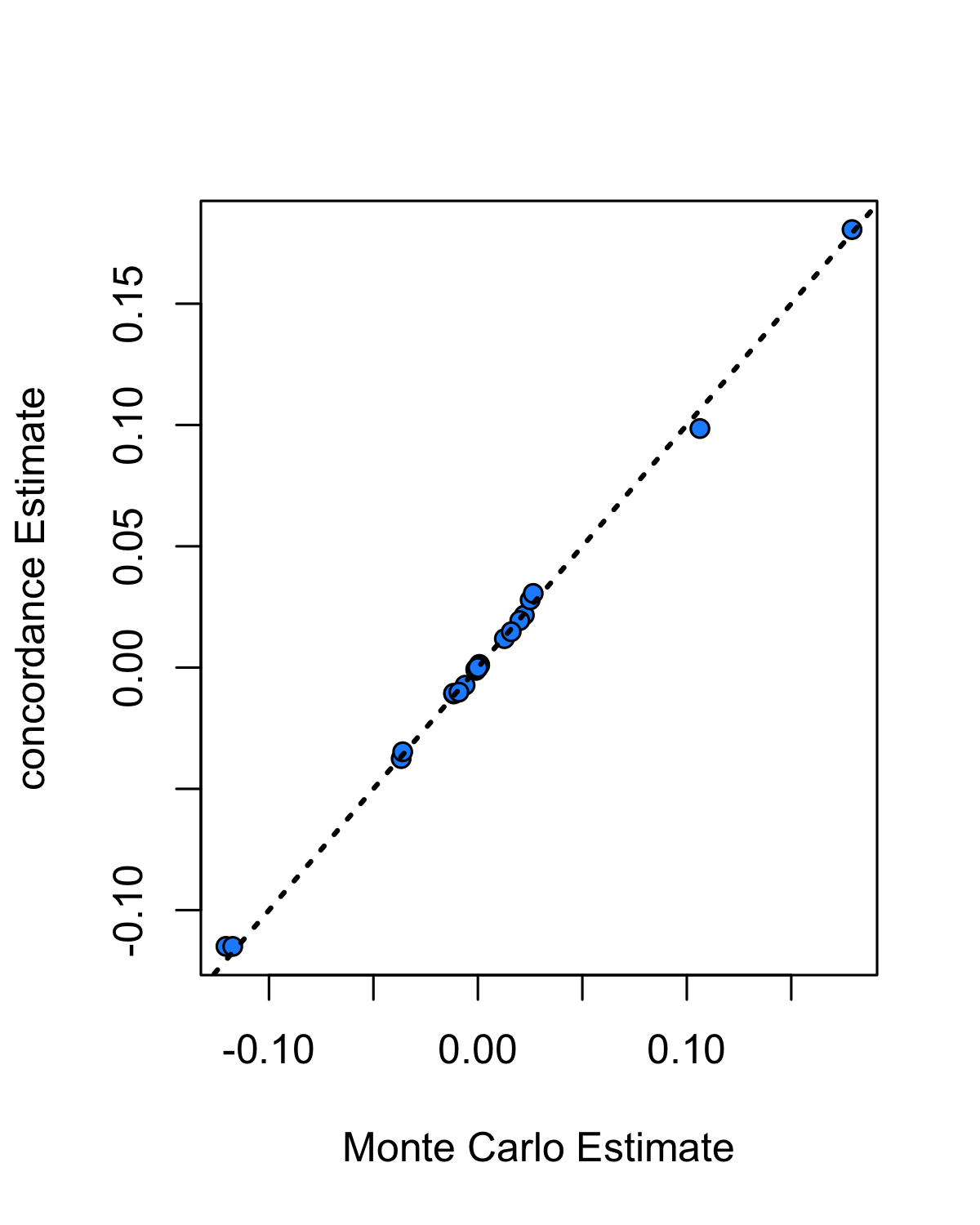}
        \caption{Comparison of $\bm C_{12}$ entries for both estimation procedures. }
    \label{fig:piston_comp}
    \end{subfigure}
    \caption{Estimation of $\bm C_{12}$ for the piston functions. }
    \label{fig:piston}
\end{figure}

\section{PBX 9501 - Concordance Values}
The following tables give the posterior mean and posterior standard deviations for the concordance analysis of the $14$ different materials for the HE application in the main text.

\begin{table}[!htbp] \centering 
\caption{Concordance posterior mean} 
  \label{} 
\begin{tabular}{@{\extracolsep{0pt}} lccccccc} 
\footnotesize
\\[-1.8ex]\hline 
\hline \\[-1.8ex] 
 & copper & tungsten & ss304 & gold & gold\_5cu & uranium\_5mo & nickel \\ 
\hline \\[-1.8ex] 
copper & $0.987$ & $0.969$ & $0.979$ & $0.915$ & $0.934$ & $0.930$ & $0.981$ \\ 
tungsten & $0.969$ & $0.998$ & $0.981$ & $0.940$ & $0.953$ & $0.942$ & $0.984$ \\ 
ss304 & $0.979$ & $0.981$ & $1.000$ & $0.930$ & $0.946$ & $0.938$ & $0.996$ \\ 
gold & $0.915$ & $0.940$ & $0.930$ & $0.991$ & $0.880$ & $0.888$ & $0.927$ \\ 
gold\_5cu & $0.934$ & $0.953$ & $0.946$ & $0.880$ & $0.999$ & $0.887$ & $0.951$ \\ 
uranium\_5mo & $0.930$ & $0.942$ & $0.938$ & $0.888$ & $0.887$ & $0.998$ & $0.941$ \\ 
nickel & $0.981$ & $0.984$ & $0.996$ & $0.927$ & $0.951$ & $0.941$ & $0.999$ \\ 
uranium\_075ti & $0.930$ & $0.946$ & $0.945$ & $0.869$ & $0.935$ & $0.944$ & $0.949$ \\ 
al6061 & $0.962$ & $0.958$ & $0.982$ & $0.906$ & $0.918$ & $0.913$ & $0.981$ \\ 
uranium & $0.899$ & $0.931$ & $0.918$ & $0.859$ & $0.954$ & $0.820$ & $0.921$ \\ 
ss250 & $0.976$ & $0.978$ & $0.995$ & $0.933$ & $0.944$ & $0.934$ & $0.992$ \\ 
tin & $0.965$ & $0.947$ & $0.972$ & $0.899$ & $0.925$ & $0.904$ & $0.973$ \\ 
ss4340 & $0.978$ & $0.980$ & $0.999$ & $0.931$ & $0.945$ & $0.937$ & $0.996$ \\ 
al7075 & $0.965$ & $0.960$ & $0.984$ & $0.907$ & $0.920$ & $0.915$ & $0.983$ \\ 
\hline \\[2ex]
 & uranium\_075ti & al6061 & uranium & ss250 & tin & ss4340 & al7075 \\ 
\hline \\[-1.8ex] 
copper & $0.930$ & $0.962$ & $0.899$ & $0.976$ & $0.965$ & $0.978$ & $0.965$ \\ 
tungsten & $0.946$ & $0.958$ & $0.931$ & $0.978$ & $0.947$ & $0.980$ & $0.960$ \\ 
ss304 & $0.945$ & $0.982$ & $0.918$ & $0.995$ & $0.972$ & $0.999$ & $0.984$ \\ 
gold & $0.869$ & $0.906$ & $0.859$ & $0.933$ & $0.899$ & $0.931$ & $0.907$ \\ 
gold\_5cu & $0.935$ & $0.918$ & $0.954$ & $0.944$ & $0.925$ & $0.945$ & $0.920$ \\ 
uranium\_5mo & $0.944$ & $0.913$ & $0.820$ & $0.934$ & $0.904$ & $0.937$ & $0.915$ \\ 
nickel & $0.949$ & $0.981$ & $0.921$ & $0.992$ & $0.973$ & $0.996$ & $0.983$ \\ 
uranium\_075ti & $0.995$ & $0.921$ & $0.885$ & $0.940$ & $0.923$ & $0.943$ & $0.922$ \\ 
al6061 & $0.921$ & $0.999$ & $0.894$ & $0.977$ & $0.964$ & $0.982$ & $0.999$ \\ 
uranium & $0.885$ & $0.894$ & $1.000$ & $0.920$ & $0.890$ & $0.918$ & $0.895$ \\ 
ss250 & $0.940$ & $0.977$ & $0.920$ & $0.999$ & $0.971$ & $0.995$ & $0.979$ \\ 
tin & $0.923$ & $0.964$ & $0.890$ & $0.971$ & $0.999$ & $0.971$ & $0.965$ \\ 
ss4340 & $0.943$ & $0.982$ & $0.918$ & $0.995$ & $0.971$ & $0.999$ & $0.984$ \\ 
al7075 & $0.922$ & $0.999$ & $0.895$ & $0.979$ & $0.965$ & $0.984$ & $1.000$ \\ 
\hline \\[-1.8ex] 
\end{tabular} 
\end{table} 

\begin{table}[!htbp] \centering 
  \caption{Concordance posterior standard deviations} 
  \label{} 
\begin{tabular}{@{\extracolsep{5pt}} lccccccc} 
\\[-1.8ex]\hline 
\hline \\[-1.8ex] 
 & copper & tungsten & ss304 & gold & gold\_5cu & uranium\_5mo & nickel \\ 
\hline \\[-1.8ex] 
copper & $0.006$ & $0.003$ & $0.004$ & $0.010$ & $0.006$ & $0.006$ & $0.004$ \\ 
tungsten & $0.003$ & $0.002$ & $0.001$ & $0.011$ & $0.005$ & $0.005$ & $0.002$ \\ 
ss304 & $0.004$ & $0.001$ & $0.0005$ & $0.010$ & $0.005$ & $0.005$ & $0.002$ \\ 
gold & $0.010$ & $0.011$ & $0.010$ & $0.011$ & $0.011$ & $0.010$ & $0.010$ \\ 
gold\_5cu & $0.006$ & $0.005$ & $0.005$ & $0.011$ & $0.002$ & $0.007$ & $0.005$ \\ 
uranium\_5mo & $0.006$ & $0.005$ & $0.005$ & $0.010$ & $0.007$ & $0.004$ & $0.006$ \\ 
nickel & $0.004$ & $0.002$ & $0.002$ & $0.010$ & $0.005$ & $0.006$ & $0.001$ \\ 
uranium\_075ti & $0.007$ & $0.006$ & $0.006$ & $0.011$ & $0.007$ & $0.009$ & $0.006$ \\ 
al6061 & $0.004$ & $0.001$ & $0.001$ & $0.010$ & $0.005$ & $0.005$ & $0.002$ \\ 
uranium & $0.005$ & $0.004$ & $0.004$ & $0.010$ & $0.007$ & $0.005$ & $0.004$ \\ 
ss250 & $0.003$ & $0.001$ & $0.001$ & $0.010$ & $0.005$ & $0.005$ & $0.002$ \\ 
tin & $0.005$ & $0.004$ & $0.004$ & $0.010$ & $0.006$ & $0.006$ & $0.004$ \\ 
ss4340 & $0.003$ & $0.001$ & $0.001$ & $0.010$ & $0.005$ & $0.005$ & $0.002$ \\ 
al7075 & $0.004$ & $0.001$ & $0.001$ & $0.010$ & $0.005$ & $0.005$ & $0.002$ \\ 
\hline \\[2ex]
 & uranium\_075ti & al6061 & uranium & ss250 & tin & ss4340 & al7075 \\ 
\hline \\[-1.8ex] 
copper & $0.007$ & $0.004$ & $0.005$ & $0.003$ & $0.005$ & $0.003$ & $0.004$ \\ 
tungsten & $0.006$ & $0.001$ & $0.004$ & $0.001$ & $0.004$ & $0.001$ & $0.001$ \\ 
ss304 & $0.006$ & $0.001$ & $0.004$ & $0.001$ & $0.004$ & $0.001$ & $0.001$ \\ 
gold & $0.011$ & $0.010$ & $0.010$ & $0.010$ & $0.010$ & $0.010$ & $0.010$ \\ 
gold\_5cu & $0.007$ & $0.005$ & $0.007$ & $0.005$ & $0.006$ & $0.005$ & $0.005$ \\ 
uranium\_5mo & $0.009$ & $0.005$ & $0.005$ & $0.005$ & $0.006$ & $0.005$ & $0.005$ \\ 
nickel & $0.006$ & $0.002$ & $0.004$ & $0.002$ & $0.004$ & $0.002$ & $0.002$ \\ 
uranium\_075ti & $0.005$ & $0.006$ & $0.006$ & $0.006$ & $0.007$ & $0.006$ & $0.006$ \\ 
al6061 & $0.006$ & $0.001$ & $0.004$ & $0.001$ & $0.003$ & $0.001$ & $0.001$ \\ 
uranium & $0.006$ & $0.004$ & $0.001$ & $0.004$ & $0.005$ & $0.004$ & $0.004$ \\ 
ss250 & $0.006$ & $0.001$ & $0.004$ & $0.001$ & $0.003$ & $0.001$ & $0.001$ \\ 
tin & $0.007$ & $0.003$ & $0.005$ & $0.003$ & $0.003$ & $0.004$ & $0.003$ \\ 
ss4340 & $0.006$ & $0.001$ & $0.004$ & $0.001$ & $0.004$ & $0.001$ & $0.001$ \\ 
al7075 & $0.006$ & $0.001$ & $0.004$ & $0.001$ & $0.003$ & $0.001$ & $0.000$ \\ 
\hline \\[-1.8ex] 
\end{tabular} 
\end{table} 

\end{document}